\documentclass[preprint,floatfix] {revtex4}

\usepackage{graphicx}
\usepackage{subfigure}
\usepackage{float}
\usepackage{xcolor}
\usepackage{enumerate}
\usepackage{braket}
\usepackage{amsmath, bm}
\usepackage{stackengine}
\usepackage{adjustbox}
\usepackage{rotating}
\usepackage{hyperref}
\usepackage{tablefootnote}

\usepackage{color, soul}
\definecolor{darkblue}{rgb}{0,0,0.5}
\setulcolor{darkblue}

\begin{document}
\title{Benchmark Computations of Nearly Degenerate Singlet and Triplet states of N-heterocyclic Chromophores : I. Wavefunction-based Methods}
\author{Shamik Chanda}
\author{Sangita Sen}
\altaffiliation{Email: sangita.sen@iiserkol.ac.in }                                           
\affiliation{Department of Chemical Sciences\\
Indian Institute of Science Education and Research (IISER) Kolkata \\  
Nadia, Mohanpur-741246, WB, India}

\begin{abstract}
In this paper we investigate the role of electron correlation in predicting the S$_1$-S$_0$ and T$_1$-S$_0$ excitation energies and hence, the singlet-triplet gap ($\Delta$E$_{ST}$) in a set of cyclazines which act as templates for potential candidates for 5th generation Organic Light Emitting Diode (OLED) materials. This issue has recently garnered much interest with the focus being on the inversion of the $\Delta$E$_{ST}$, although experiments have indicated near degenerate levels with both positive and negative being within the experimental error bar (\emph{J. Am. Chem. Soc. 1980, 102: 6068 , J. Am. Chem. Soc. 1986, 108: 17} ). We have carried out a systematic and exhaustive study of various excited state electronic structure methodologies and identified the strengths and shortcomings of the various approaches and approximations in view of this challenging case. We have found that near degeneracy can be achieved either with a proper balance of static and dynamic correlation in multireference theories or with state-specific orbital corrections including its coupling with correlation. The role of spin contamination is also discussed. Eventually, this paper seeks to produce benchmark numbers for establishing cheaper theories which can then be used for screening derivatives of these templates with desirable optical and structural properties. Additionally we would like to point out that the use of DLPNO-STEOM-CCSD as the benchmark for $\Delta$E$_{ST}$ (as used in \emph{J. Phys. Chem. A 2022, 126: 8: 1378, Chem. Phys. Lett. 2021, 779: 138827}) is not a suitable benchmark for this class of molecules.


\end{abstract}
\maketitle   

\section{Introduction} \label{intro}
A new generation of materials have been proposed for making Organic Light Emitting Diodes (OLEDs) which can potentially have 100\% quantum yield\cite{jacs80,jacs86}. The design principle involves having near-degenerate S$_1$ and T$_1$ states such that the electrons that have undergone intersystem-crossing (ISC) from S$_1$ to T$_1$ can be brought back to S$_1$ by reverse intersystem crossing (RISC) through thermal activation and subsequent deexcitation occurs from the S$_1$ to S$_0$ state, which leads to thermally activated delayed fluorescence (TADF)\cite{endo2011,uoyama2012,desilva2019a}. It is reasonable to suppose that RISC would be most efficient if no thermal activation was required. This can be achieved if the S$_1$-T$_1$ gap is inverted\cite{difley2008,sato2015,di2017,olivier2018}. However, one must not forget that the mechanism of RISC involves vibronic coupling which diminishes with increasing energy gap\cite{dinkelbach2021}. Thus, theoretical screening of OLED candidate molecules for large inverted gaps may be counterproductive. In fact, the experimental values for the few candidate molecules which could be synthesized show small positive/negative values\cite{nature2022}. Benchmarking cheaper, mostly density functional approaches, against theories indicating large inverted gaps can thus be misleading. Only static correlation, on the other hand, is very sensitive to the choice of the active space and thus tedious for large scale screening. Supplementing CASSCF with dynamic correlation can often reduce this sensitivity. In this paper we examine the correlation effects captured by various wavefunction-based electronic structure theories and try to arrive at the most accurate correlation model to serve as a benchmark for DFT-based methods. A few preliminary DFT-based computations are presented here while a more detailed study follows in a forthcoming publication\cite{saha2023}.

\begin{figure}[h]
\centering

  \centering
	\includegraphics[width=0.6\textwidth]{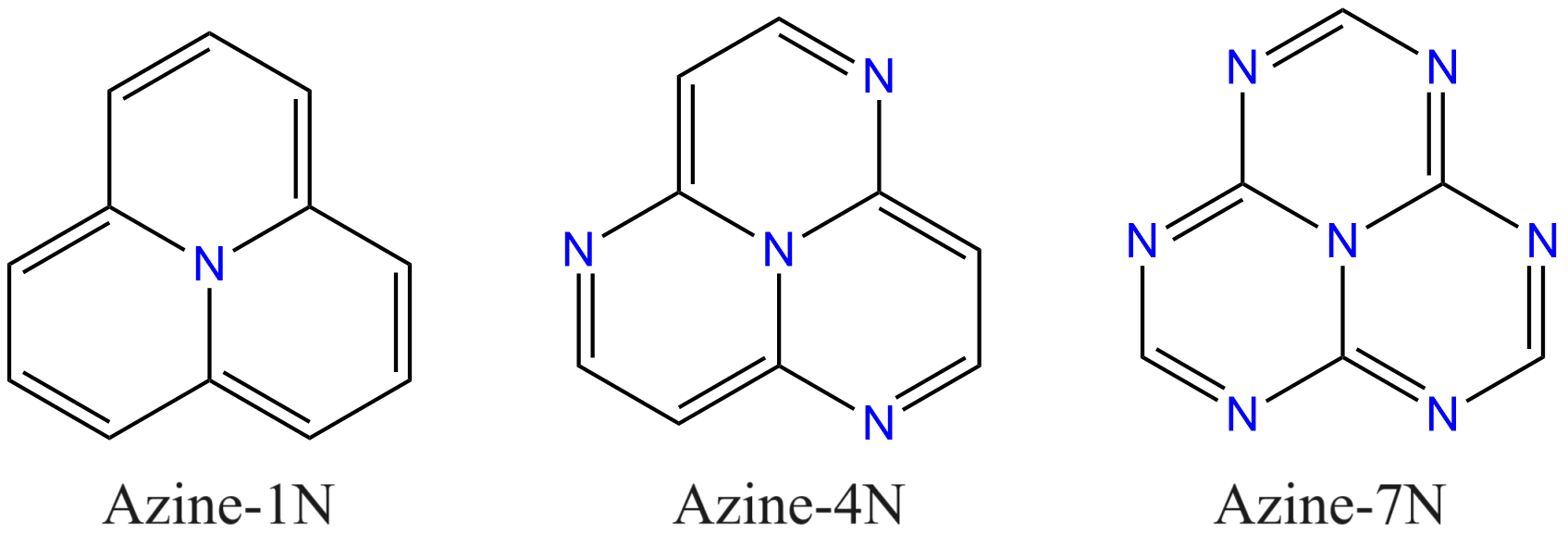}
  \caption{The molecular templates studied in this paper for inverted singlet-triplet gaps.}
  \label{molecule}

\end{figure}

\begin{figure}[h]
\centering

  \centering
	\includegraphics[width=0.5\textwidth]{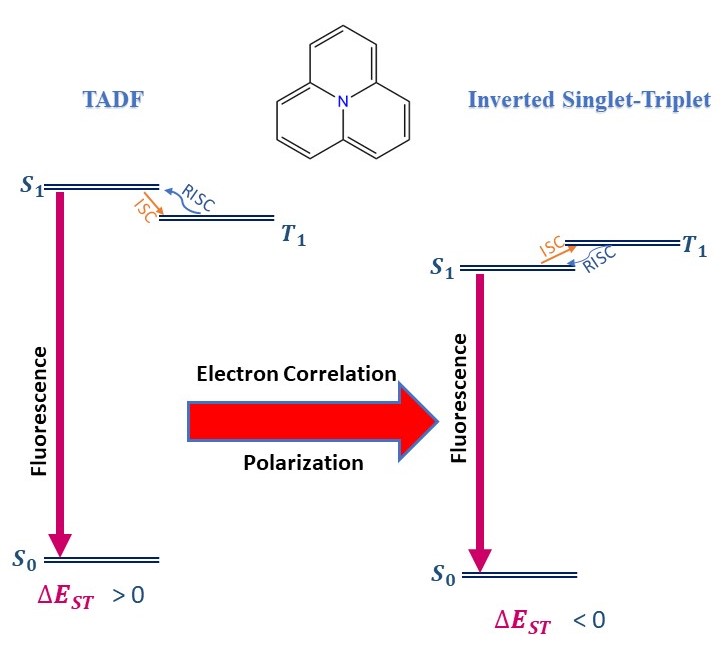}
  \caption{Violation of Hund's Law}
  \label{rule_violate}

\end{figure}

Exchange interactions, which stabilise the T$_1$ state result in T$_1$ to be typically below the S$_1$ state with $\Delta E_{ST} > 0$ (Hund's Law\cite{hund1925}). However, some N-substituted fused ring molecules, such as cyclazines and their derivatives (see Fig. \ref{molecule}) exhibit close to degenerate or inverted S$_1$-T$_1$ gaps in experiments\cite{jacs80,jacs86,nature2022}. Common linear response excited state methods such as Configuration Interaction Singles (CIS)\cite{pople2003,foresman1992}, Random Phase Approximation (RPA)\cite{bouman1983,bouman1989} or Time-Dependent Density Functional Theory (TD-DFT)\cite{casida1995} which are formulated for treating primarily singly excited states\cite{dreuw2005} almost always give this gap as positive. While CIS and RPA are theoretically guaranteed to give $\Delta$E$_{ST}>0$, TD-DFT can give an inverted gap if the exchange-correlation functional captures sufficient electron correlation which is usually not the case for commonly used functionals.  In recent publications\cite{desilva2019b,ghosh2022,ricci2021}, presence of low-lying excited states with doubly excited character have been indicated to be the crucial criteria for ST inversion\cite{desilva2019b}. Simply put, greater electron correlations in S$_1$ relative to T$_1$ reduce the ST gap leading to a nearly degenerate or even inverted $\Delta$E$_{ST}$ as shown in Fig.~\ref{rule_violate}.

Recently, in this context, azine-1N, azine-4N, azine-7N (heptazine) (see Fig.~\ref{molecule}) and related molecules and derivatives have received much interest from the electronic structure community\cite{ghosh2022,desilva2019b,ricci2021,li2022,police2021,nature2022,domcke2021,dreuw2023,rodrigo2021,ehrmaier2019,pios2021,tuckova2022,loos2023}. The inability of linear-response time-dependent density functional theory (LR-TDDFT)\cite{runge1984,casida1995} and the success of wave function techniques such as doubles-corrected configuration interaction singles [CIS(D)]\cite{rhee2007}, and equation of motion coupled cluster singles and doubles (EOM-CCSD)\cite{headgordon1994,stanton1993} in computing inverted singlet-triplet gaps have been pointed out\cite{desilva2019b,ghosh2022,sanchogarchia2022}. A combined DFT/Multireference Configuration Interaction (DFT/MRCI) method to investigate the impact of the negative singlet-triplet gap and the vibronic coupling in heptazine derivatives was also recently reported\cite{dinkelbach2021}. Novel techniques such as spin-component scaled second-order coupled cluster (SCS-CC2)\cite{hellweg2008}, ADC\cite{wormit2014}, and multireference techniques such as complete active space self-consistent field (CASSCF)\cite{olsen1988} and N-electron valence second-order perturbation theory (NEVPT2)\cite{angeli2001} have also predicted the inverted singlet-triplet gaps\cite{ricci2021,rodrigo2021}. Recently Loos. et. al.\cite{loos2023} reported benchmark excitation energies for a set of azine and heptazine based molecules using a high level CC3\cite{christiansen1995} framework. Computationally cheaper domain-based local pair natural orbital (DLPNO) similarity transformed EOM-CCSD (STEOM-CCSD)\cite{nooijen1997} also inverted the gaps but unfortunately they are far more inverted than predicted by experiments(\cite{jacs80,jacs86}) and inconsistent across various molecules\cite{ghosh2018,ghosh2022}.

Among the density functional approaches, the doubles-corrected TDDFT [TDDFT(D)]\cite{grimme2007} approach with B2PLYP double-hybrid functionals, could achieve inversion\cite{police2021} but since the DLPNO-STEOM-CCSD was considered as the benchmark, they were dismissed as having inadequate correlation. In this (see Fig.~\ref{1N,4N,7N}), and a forthcoming publication\cite{saha2023} we demonstrate that these functionals are indeed reasonable and it is the DLPNO-STEOM-CCSD which over-inverts. This is an important finding as the TD-DFT framework is very popular for screening molecules for stable molecular structures, peak wavelengths, intensities, colour purity and other desirable structural and optical features. A proper choice or design of a DFT framework is predicated by the availability of a benchmark theory. CC2 and ADC(2) are also possible alternatives with a similar computational cost\cite{ricci2021,loos2023}. In this particular case, the canonical EOM-CCSD may be used but we cannot take advantage of the cost-saving afforded by the similarity transformed EOM (STEOM) approach (see Table~\ref{tab:singref}). The aim of our paper is, thus, to suggest the most reliable benchmark theory for this class of molecules and in doing so, to properly establish the necessary physics that needs to be captured to be able to describe the energetics of the optical processes involved here.

A brief comparison of single and multi-reference theories used in this article is given in Sec.~\ref{theory}. The computational and technical details is given in Sec~\ref{comput}. In Sec~\ref{singref}, a detailed analysis of the results obtained from single-reference theories are given including CIS, CIS(D), TDHF/RPA, RPA(D)\cite{haase2020}, ADC(2) and EOM-CCSD. Then, in Sec~\ref{multiref}, the detailed analysis of the CASSCF for several choices of active electrons and orbitals is carried out for a prototypical molecule, ie. azine-1N. This section also includes the CASSCF, strongly contracted(SC)-NEVPT2, CASPT2, and FICMRCI results for all the molecules. Finally we summarize our conclusions and discuss the future outlook of this study in Sec~\ref{summary}.

\section{Theoretical Details}\label{theory}

Any N-electron wavefunction can be completely described in a given M-orbital basis as a linear combination of all possible Nth order determinants that can be constructed from the M orbitals\cite{szabo96}. However, for the ease of approximation a distinction is made between two subsets of determinants. In a typical electronic state, it is found that a few determinants constructed usually from degenerate or quasi degenerate orbitals have significantly higher coefficients. These determinants are said to constitute static/non-dynamic correlation. For wavefunctions which are eigenstates of the non-relativistic Hamiltonian, one can further distinguish linear combinations necessary to spin-adapt the N-electron function to a given $<S^2>$ value and those that are necessary for accuracy. A pre-adaptation of the reference functions to a given $<S^2>$ value by forming configuration state functions (CSFs) is also possible - in which case the CSFs other than the reference CSF would be said to constitute static correlation. On the other hand, dynamic correlation energy is the energy contribution of the large number of remaining determinants with small coefficients. The collective contribution of these determinants is often too significant to ignore. In our case, we shall see that both static and dynamic correlations are important.

The simplest ab initio electronic structure theory for excited states is the Configuration Interaction Singles (CIS)\cite{foresman1992} theory. 
The lowest singlet and triplet excited state CSF's, formed by exciting an electron from orbtal $i\to a$, in terms of Slater determinants $|\psi_i ^a \rangle$, can be written as, 
\begin{align*}\tag{1}
    |^1 \phi_i ^a \rangle = \frac{1}{\sqrt 2}(|\psi_i ^a \rangle+|\psi_{\bar{i}}^{\bar{a}}\rangle) \\
    |^3 \phi_i ^a \rangle =\frac{1}{\sqrt 2}(|\psi_i ^a\rangle-|\psi_{\bar{i}} ^{\bar{a}}\rangle )
\end{align*}
The lowest eigenvalue of the difference of the S$_1$ and T$_1$ Hamiltonian matrices can be shown to be the lowest singlet-triplet gap $\Delta E_{ST}$ using Weyl's inequality in the following form\cite{desilva2019b},
\begin{equation*}\tag{2}
    \Delta E_{ST} = E_{S_1}-E_{T_1}=\lambda_{min}(^1H)+\lambda_{max}(-^3H) \ge \lambda_{min}(^1H-^3H)
\end{equation*}
where, 
\begin{equation*}\tag{3}
(^1H-^3H)_{ia,jb}=\langle^1\phi_i ^a|H|^1\phi_j ^b\rangle - \langle^3\phi_i ^a|H|^3\phi_j ^b\rangle = 2\langle \psi_i ^a |H| \psi_{\bar{j}} ^{\bar{b}} \rangle = 2(ia||\bar{jb})=2(ia|jb)
\end{equation*}
This is simply twice of the conventional, antisymmetrized, two-electron exchange integrals in chemists' notation, which is known to be positive semidefinite\cite{helgaker2000}. Following the arguments in\cite{desilva2019b}, its lowest eigenvalue, $\lambda_{min}(^1H-^3H) \ge 0$ and $\Delta$E$_{ST} \ge 0$ for any CIS Hamiltonian. 
This implies that, within a single excitation framework (for an uncorrelated theory), this gap is always positive leading to a positive ST gap. It is also easy to see from here that since electron correlations typically lower the energies of electronic states and relatively higher amount of correlation in S$_1$ as compared to T$_1$ could, in principle, make the ST gap negative. In TD-DFT\cite{casida1995,gross1984,burke2005}, electron correlations are accounted for through the exchange–correlation kernel, $f_{xc}(r,r',\omega$)\cite{gross1985}. By analogy, we can also write a similar equation for TD-DFT (within the Tamn-Dancoff approximation), 
\begin{equation*}\tag{4}
    \langle^1\phi_i ^a|H|^1\phi_j ^b\rangle - \langle^3\phi_i ^a|H|^3\phi_j ^b\rangle=2(ia|jb)+ 2(ia|f_{xc}^{\alpha,\beta} (r,r',\omega)|jb) 
\end{equation*}
The exchange-correlation kernel $f_{xc}(r,r',\omega)$ is negative definite\cite{gross1985,burke2012,dobson2000} leading to a reduction of the ST gap and ideally to inversion. But, for commonly known functionals this gap is always $>0$ as sufficient correlation is not captured. It is well-known that TD-DFT fails to describe double excitations if a frequency-independent kernel is used (ie. under the adiabatic approximation)\cite{elliott2011,burke2005}. On the other hand, double hybrid functionals\cite{grimme2007} introduce double excitations through a post-SCF MP2 correction and can give an inverted ST gap\cite{police2021,sanchogarchia2022,ghosh2022}. We justify this in our forthcoming publication and compare various functionals through proper benchmarking\cite{saha2023}. An alternative path to obtain open-shell singlet excited states at the SCF level is to use a modified algorithm to converge to a higher state in a brute-force manner leading to the $\Delta$SCF methods. There are several strategies for this with the most common ones being the maximum overlap method (MOM)\cite{gilbert2008}, square-gradient minimization method (SGM)\cite{hait2020,hait2021}, restricted open-shell Kohn-Sham (ROKS)\cite{hait2016}, excited state DFT (e-DFT)\cite{edft2008} etc. However, the $\Delta$SCF methods suffer from severe spin contamination mostly due to forceful representation of a multi-determinant wave-function as a single-determinant. Non-orthogonality of ground and excited state functions makes it very difficult to obtain transition properties. Thus, we believe that $\Delta$SCF methods while cheap and convenient should be restricted to initial screening applications and not for detailed study. Correlation methods such as MP2 and CCSD can also be applied on the optimized HF excited states leading to $\Delta$MP2 and $\Delta$CCSD excitation energies.

CIS(D) includes a perturbative doubles correction on the CIS excitation energies akin to MP2 giving size consistent energies as opposed to CISD. EOM-CCSD may be considered as the linear response of the CCSD ground state function. The EOM-CCSD excitation energies show an exact cancellation of the common correlation terms between the ground and excited states. The differential orbital relaxation is included up to linear terms but the differential correlation is not included under the SD truncation.

The most common starting point for a multireference theory\cite{lischka2018} is the Multi-Configuration SCF (MCSCF). In the MCSCF approach, an active N-electron space needs to be defined. The selection is based on choosing a set of quasi-degenerate molecular orbitals or those that are the most relevant in describing the chemical bonding or the electronic excitations of interest. If the active electrons are distributed in all possible ways to make the model functions, a Complete Active Space (CAS) is created. The CASSCF\cite{olsen1988} function for the k$^{th}$ state is then written as:
\begin{equation*}\tag{5}
    \ket{\psi_k} = \sum_{\mu=1}^{n_{det}} \ket{\phi_\mu}c_{\mu k} = \sum_{\mu=1}^{n_{CSF}} \ket{\Phi_\mu}d_{\mu k} \forall \ k
\end{equation*}
The Complete active space (CAS) wavefunction is thus expressed as a linear combination of singly-, doubly-, triply-substituted, etc. Slater determinants, but with the excitation operator confined within the subset of the selected active orbitals. 
\begin{equation*}\tag{6}
    |\Psi_{CAS}\rangle = \Sigma_{M} |\Psi_{M}\rangle = C_0 |\Psi_0\rangle +\Sigma_{i,a} C_{i} ^{a} |\Psi_i ^a\rangle + \Sigma_{ij,ab} C_{ij} ^{ab} |\Psi_{ij} ^{ab}\rangle + \Sigma_{ijk,abc} C_{ijk} ^{abc} |\Psi_{ijk} ^{abc}\rangle + ..... 
\end{equation*}
Here $C_{i} ^{a}$, $C_{ij} ^{ab}$, $C_{ijk} ^{abc}$, etc. are the corresponding coefficients of expansion, indicating singly, doubly, triply, etc. excited configurations respectively. The relative weights of the sums of square of $C_{ij} ^{ab}$, $C_{ijk} ^{abc}$ , etc. coefficients reflects the importance of doubly, triply n-tuply excited states beyond the singly excited ones.  
Doubly excited configurations are thus explicitly present in the excited-state wave function and excited states with doubles character can be obtained with equal facility as those with dominance of singles unlike CIS, RPA and TD-DFT.
The primary objective of CASSCF is to accurately describe the static electronic correlation. By allowing a flexible choice of active space, CASSCF can handle systems with significant static correlation, such as molecules in the bond-breaking or bond-forming regions as well as highly multireference excited states.

Inclusion of further dynamic correlation on top of the multireference function, can be carried out through perturbative\cite{anderson1990,Andersson1992SecondorderFunction,Hirao1992MultireferenceMethod,Kozlowski1994ConsiderationsTheory,Hoffmann1996CanonicalVariables,angeli2001,Sen2015UnitaryApplications}, CI-like\cite{Werner1988AnIN} or CC-like\cite{Jeziorski1981,Li1995UnitaryApproximations,Li1997ReducedStates,MAHAPATRA1998AApplications} approaches. These constitute the family of multireference correlation theories which are called uncontracted if the ansatz for dynamic correlation excites out of the individual $\phi_\mu$ and contracted if it excites directly out of the $\psi_k$\cite{MAHAPATRA1998163}. In this paper we have selected Strongly-contracted N-Electron Valence Perturbation Theory Second Order (SC-NEVPT2)\cite{angeli2001}, Complete Active Space Perturbation Theory Second Order (CASPT2)\cite{anderson1990,Andersson1992SecondorderFunction} and Fully Internally Contracted Multireference Configuration interaction singles-doubles (FIC-MRCISD)\cite{Werner1988AnIN} as representative theories from the various classes of multireference correlation theories. The Internally Contracted Multireference Coupled Cluster (ICMRCC)\cite{aoto2016} theory turned out to be too computationally expensive for these molecules in the chosen basis. All three belong to the contracted category.

NEVPT2\cite{angeli2001} accounts for the dynamic correlation effects by considering the interaction between the active space and the remaining inactive orbitals and electrons in a perturbative manner.  It is particularly useful for treating systems with moderately high dynamic correlation effects. FIC-MRCISD\cite{Werner1988AnIN} (denoted as FIC-MRCI in this paper) is a non-perturbative diagonalization method that includes a larger number of configurations compared to NEVPT2 and a complete linear coupling between then, allowing it to capture a more extensive range of dynamic correlation effects. The FIC-MRCI is the best possible benchmark method for medium-sized highly correlated molecular systems such as in this current article and we shall use it as our benchmark theoretical method. It has predicted excitation energies and ST gaps consistent with experiments, which aligns with our theoretical expectations. The issue of size extensivity\cite{Szalay1995}, which is lacking in FIC-MRCI, can be overlooked for excitation energy computations but for bond-breaking situations it becomes critical, and NEVPT2 and ICMRCC\cite{aoto2016}, which are fully size extensive would then be more reliable.

In the next sections we compare and analyze the performance of these theories in the context of predicting the S$_1$-S$_0$ and T$_1$-S$_0$ excitation energies and hence, the singlet-triplet gap ($\Delta$E$_{ST}$) in a set of cyclazine-based templates shown in Fig.~\ref{molecule}.

\section{Computational Details}\label{comput}
All CIS, CIS(D), RPA, RPA(D) and DLPNO-STEOM-CCSD calculations were performed using the ORCA 5.2.1 package\cite{orca2020}. All the EOM-CCSD, ADC(2) and $\Delta$CCSD calculations using the Maximum Overlap Method (MOM)\cite{gilbert2008} are performed using a trial version of the Q-Chem 6.0.1 package\cite{shao2015}. The state average complete active space self consistent field (SA-CASSCF) calculations along with FIC-CASPT2, SC-NEVPT2 and FIC-MRCI are done in ORCA 5.2.1\cite{orca2020}. For the SA-CASSCF calculations two singlet states S$_0$ and S$_1$ and one triplet state T$_1$ has been averaged out.
Alrich's def2-TZVP basis set\cite{weigend2005} has been used unless explicitly mentioned. For the $\Delta$CCSD(T)\cite{Stanton1997,Raghavachari1989} calculations, the cc-pVDZ\cite{dunning1989} basis set has been employed.   
The molecular geometries are optimized in ORCA using B3LYP\cite{becke1993}/def2-TZVP with Grimme's $D_3$ dispersion corrections\cite{grimme2011}. Vibrational frequency analyses were performed on all optimized geometries to confirm their nature as global minima. A tight SCF convergence criterion ($10^{-8}$) was fixed for all calculations. For a few cases among the $\Delta$SCF methods, it was difficult to converge the S$_1$ state with the MOM strategy. In these cases we relaxed the convergence criterion to $10^{-6}$. All the absolute values of state energies in atomic units ($au$) are reported in the Table.~S1A and S1B of the supplementary material.
The FIC-MRCISD and EOM-CCSD computations were the most computationally expensive methods with runtime on a single CPU of the order of 10-14 hours while all the DFT methods took less than half an hour for each run. We emphasize that the purpose of the state-of-the-art computations is to provide benchmark numbers for proper evaluation of cheaper methods and identification of necessary physics.

\begin{figure}[h]
\centering

  \centering
	\includegraphics[width=1\textwidth]{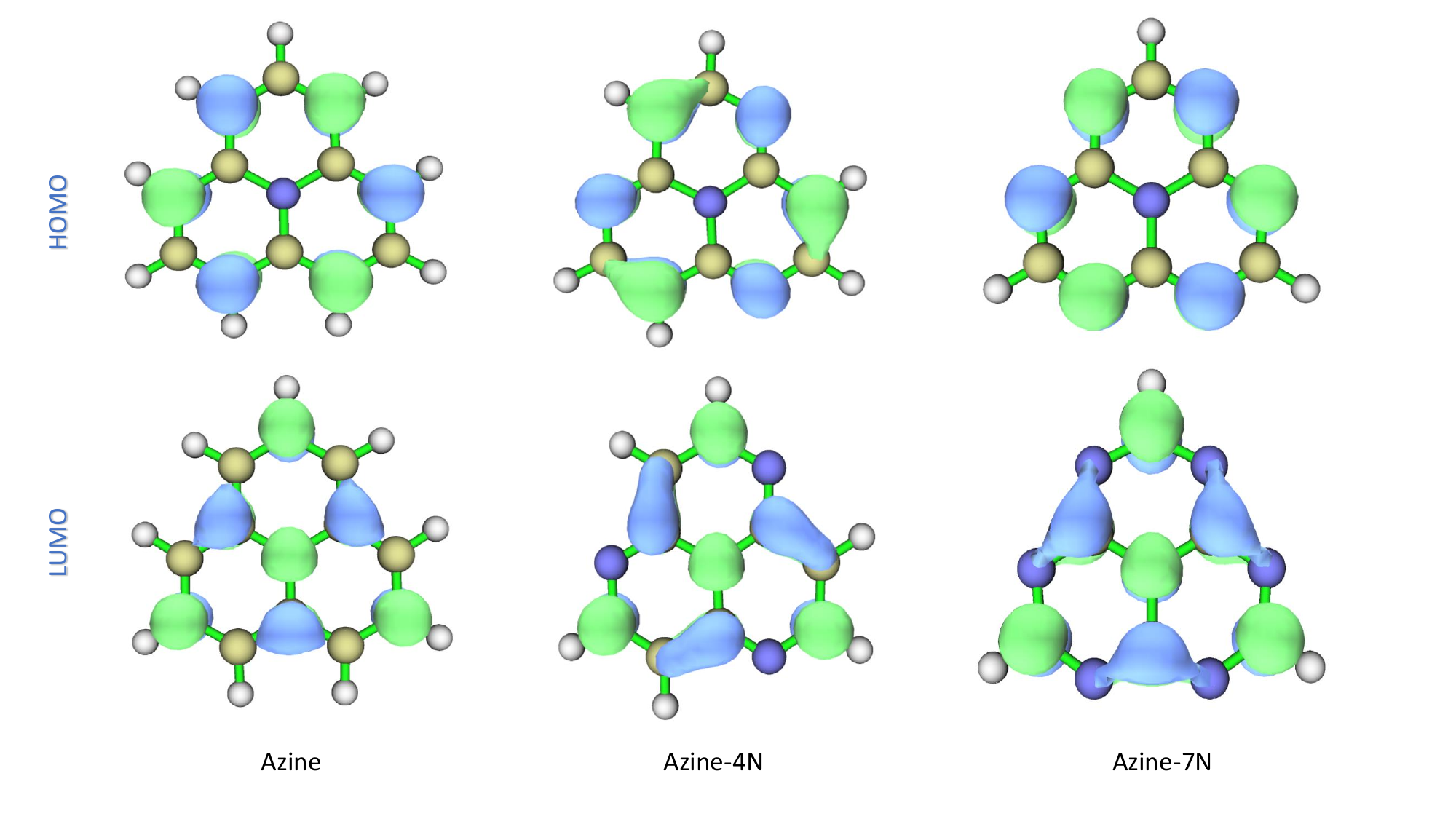}
  \caption{HOMO-LUMO of three molecules in Fig.~1 with B3LYP/def2-TZVP level of theory}
  \label{HOMO-LUMO}

\end{figure}

\section{Results and Discussions}\label{results}
In this article, we have looked at the electronic structures of a set of three molecules (given in Fig~1) in their ground, S$_1$, and T$_1$ states.  Azine-1N and azine-4N are two of the very first compounds that were found to have inverted singlet-triplet gaps. The experiment determined the singlet-triplet gaps for azine-1N and azine-4N to be -0.07\cite{jacs80} and $<0.1$\cite{jacs86} eV, respectively. Recent theoretical investigations\cite{desilva2019b,police2021,bhattacharyya2021,ghosh2022,ricci2021,domcke2021} have also reported inverted ST gaps. Azine-7N or heptazine has been extensively researched in recent years for its theoretically predicted large inverted singlet-triplet gap of about -0.25 eV\cite{desilva2019b}.
The HOMO and LUMO of these molecules are given in Fig~3, with clearly visible spatially separated electron densities. This reveals the charge transfer (CT) character of the excited states in all three molecules\cite{desilva2019b} and indicates a suitability for OLED applications which has further fueled interest in these molecules. We can observe a reduction in radicaloid nature of the ground-state (S$_0$), from azine-1N to azine-7N, as indicated by the integrated number of electrons ($N$) housed on fractionally occupied orbitals (see Table~S2 from supplementary material).

In the discussions that follow, we divide our work into two groups - single reference and multi-reference based methods. This helps us to disentangle the role of dynamic correlation and static correlation. It is most important to realize that the terminology of `doubles', `triples', etc. needs a qualifier indicating the reference function. Since the S$_1$ and T$_1$ states are primarily singly excited with respect to S$_0$, many of the `doubles' with respect to S$_0$ (which are the most important contributors to the dynamic correlation of S$_0$) are only `singles' with respect to S$_1$ or T$_1$. Conversely, the `doubles' necessary to correlate S$_1$ and T$_1$ are `triples' relative to S$_0$. As the various correlated electronic structure methods include varying amounts of these excited functions, their suitability for different types of electronic states differ. From a different perspective, the successes and failures of different methodologies gives us an insight into the nature of the electronic states under study. We also include a few selected DFT-based methods to emphasize our conclusions in this paper and present a thorough benchmark study in a paper to follow\cite{saha2023}.

For these azine-xN (x=1,4,7) molecules, the most consistent description of their correlation that we have arrived at from a careful study of the comparative numbers is as follows. The S$_0$ state is single-reference but has a large amount of dynamic correlation. The S$_1$ state is multi-reference and is dominated by a singly excited configuration out of S$_0$ but contains significant contributions from doubles ($\approx$7\% for azine-1N) and triples ($\approx$1-2\% for azine-1N). The S$_1$ state has a large amount of dynamic correlation too but most of it comes from configurations common between the correlated S$_0$ and S$_1$ functions. The T$_1$ state, on the other hand, is also primarily singly excited with significant contributions from doubles ($\approx 5-6\%$ for azine-1N) but a larger triples contribution ($\approx 2-3 \% for azine-1N$) than S$_1$. The dynamic correlation of T$_1$ come from configurations not common to those correlating S$_0$. The most important correlation effect is the spin-dependent (or equivalently, state-specific) orbital relaxation in the presence of correlation. This is sometimes also called spin-polarization in more general terms. 
We justify the above assertions in the discussions that follow. In earlier publications, the dynamic correlation of S$_1$ has been the main focus and little attention has been paid to T$_1$\cite{ghosh2022,desilva2019b,ricci2021}. However, we have found in the course of our investigations that the S$_1$-S$_0$ gap is far easier to model than the T$_1$-S$_0$ gap. This confusion stems partly from the desire to obtain a large inversion of the S$_1$ and T$_1$ which comes about when S$_1$ is properly correlated, but T$_1$ is not, leading to a premature celebration of the success of the theories which invert the gap. A proper balanced treatment leads to near degeneracy in line with experiments. A similar conclusion has been reached by Dreuw and Hoffmann \cite{dreuw2023} both with respect to including more correlation as well as larger basis sets. They have also analyzed the nature of the S$_1$ and T$_1$ states in terms of entanglement entropy and the so-called natural transition orbital participation ratio.

\subsection{Single-reference framework} \label{singref}

\begingroup                      
\begin{table}
\begin{ruledtabular}
\caption{\label{tab:singref}Vertical excitation energies (in eV) using single-reference methods. Geometries are optimized at B3LYP/def2-TZVP level.}
\vspace{0.1in}
\resizebox{0.75\textwidth}{!}{
\begin{tabular}{ c  c  c  c  c}
Molecule& Methods &	S$_1$&	T$_1$&	$\Delta$E$_{ST}$\\
\hline
Azine-1N  & TDHF/RPA/def2-TZVP & 1.639 & 0.993 &  0.646 \\     
    & RPA(D)/def2-TZVP   & 0.909 & 1.503 &  -0.59 \\
    & CIS/def2-TZVP      & 1.791 & 1.451 &  0.340 \\
    & CIS(D)/def2-TZVP   & 1.036 & 1.317 &  -0.280 \\
    & EOM-CCSD/def2-TZVP & 1.068 & 1.146 & -0.077 \\
    & DLPNO-STEOM-CCSD/def2-TZVP & 0.689 & 1.113 & -0.42 \\
    & EOM-CCSD/cc-pVDZ & 1.094 & 1.201 & -0.107 \\
    & STEOM-CCSD/cc-pVDZ & 0.592 & 1.091 & -0.499 \\
    & DLPNO-STEOM-CCSD/cc-pVDZ & 0.631 & 1.118 & -0.487 \\
    & ADC(2)/def2-TZVP   & 1.036 & 1.215 &  -0.17 \\
    & ADC(2)/cc-pVTZ$^a$ & 1.02 & 1.16 & -0.14 \\
    & ADC(3)/cc-pVTZ$^a$ & 0.81 & 0.87 & -0.06 \\
    & CC3/aug-cc-pVTZ$^d$ & 0.979 & 1.110 & -0.131 \\
   & $\Delta$HF/cc-pVDZ & 1.012 & -0.763 & -1.775\\
   & $\Delta$MP2/cc-pVDZ & 1.434 & 2.875 & 1.441 \\
   & $\Delta$CCSD/cc-pVDZ & 0.881 & 1.190 & -0.309 \\
   &$\Delta$CCSD(T)/cc-pVDZ & 1.182 & 1.187 & -0.005 \\
   & $\Delta$HF/def2-TZVP & 0.955 & -0.028 & -0.983 \\
   & $\Delta$MP2/def2-TZVP & 1.423 & 2.958 & 1.535 \\
   & $\Delta$CCSD/def2-TZVP & 0.834 & 1.139 & -0.305 \\
   & $\Delta$CCSD(T)/cc-pVTZ$^a$ & -& - &  0.025\\
    & $\Delta$CCSD(T)/aug-cc-pVTZ$^a$ & -& - &  0.029\\
    &Expt$^b$ & 0.97 & 0.93/1.05 & 0.04/-0.07\\
\hline				
Azine-4N &      TDHF/RPA/def2-TZVP  &     3.163 & 1.188 & 1.975 \\ 
 &      RPA(D)/def2-TZVP	  &     2.141 & 3.872 & -1.731 \\
 &      CIS/def2-TZVP	  &     3.297 & 2.526 & 0.771 \\
 &      CIS(D)/def2-TZVP	  &     2.226 & 2.474 & -0.248 \\
 &      EOM-CCSD/def2-TZVP  &     2.382 &	2.213 &	0.169 \\
 &      DLPNO-STEOM-CCSD/def2-TZVP  &     1.889 & 2.082 & -0.193 \\
 &      ADC(2)/def2-TZVP	  &     2.147 &	2.138 &	0.009 \\
 &      Expt$^c$      &  $<$2.39 & 2.29 & $<$0.1\\
\hline				
Azine-7N &    TDHF/RPA/def2-TZVP   &    4.243 & 3.671 & 0.572  \\
   &    RPA(D)/def2-TZVP	   &    2.494 & 3.104 & -0.61 \\
   &    CIS/def2-TZVP	   &    4.351 & 3.94  & 0.411 \\
   &    CIS(D)/def2-TZVP	   &    2.649 & 3.172 & -0.523 \\
   & EOM-CCSD/def2-TZVP & 2.916 & 3.066 & -0.150 \\
   &    DLPNO-STEOM-CCSD/def2-TZVP   &    2.297 & 3.531 & -1.234 \\
   &    ADC(2)/def2-TZVP  &  2.668029 & 2.916 & -0.248 \\
   & ADC(3)/cc-pVTZ$^a$ & 2.81 & 2.88 & -0.07 \\
   & CC3/aug-cc-pVTZ$^d$ & 2.717 & 2.936 & -0.219 \\
   &    Expt$^c$       & $--$ & $--$ & $<$0\\
\end{tabular}
}
\end{ruledtabular}
\begin{tabbing}
$^{a}$RI-MP2/cc-pVTZ optimized geometry\cite{dreuw2023}.
$^{b}$Ref\cite{jacs80}.
$^{c}$Ref\cite{jacs86}.
$^{d}$Ref\cite{loos2023}.

\end{tabbing}
\end{table}
\endgroup

All the results from single reference theories are presented in Table~1. As explained in Sec.~\ref{theory}, in uncorrelated theories such as  CIS and TD-HF, the ST gap is predicted to be positive. 
When doubly excited configurations are included in any way, the situation is drastically altered. The gap is predicted to be negative by the various correlated wave function-based approaches, CIS(D), ADC(2), EOM-CCSD, DLPNO-STEOM-CCSD and RPA(D) although they do not all quantitatively agree. CIS(D), ADC(2) and EOM-CCSD approach the experimental S$_1$-S$_0$ gap (and also our theoretical benchmark FIC-MRCI (12,9)) to within approximately 0.1~eV for all three molecules) but the T$_1$-S$_0$ gap turns out to be more challenging. RPA(D) and CIS(D) give a too high value but a full canonical EOM-CCSD (1.146 eV for azine-1N) is reasonable. Our comparison between EOM-CCSD and STEOM-CCSD for azine-1N shows that the similarity transformed EOM (STEOM) approximation is not reliable for this type of molecules. We must point out that the DLPNO-STEOM-CCSD method has been used as a benchmark in certain recent publications leading to confusion regarding the relative accuracy of various methods\cite{ghosh2022,bhattacharyya2021}. Coming to the ADC-based methods, ADC(2) performs reasonably well in the def2-TZVP basis chosen by us but a comparison between ADC(2)/cc-pVTZ and ADC(3)/cc-pVTZ from Ref\cite{dreuw2023} shows a deterioration of individual excitation energy values although the accuracy of the gap is maintained by error cancellation. The $\Delta$CCSD and $\Delta$CCSD(T) numbers adapted from Ref\cite{dreuw2023} and our own computations in the cc-pVDZ basis set show that both energy gaps are poorly described by $\Delta$CCSD relative to EOM-CCSD but the triples (T) push them in the right direction again. We justify all these observations in the next few paragraphs.

CIS(D) and EOM-CCSD include a set of doubles on S$_0$. The contribution of these doubles is estimated in an MP2-like state-specific fashion for CIS(D) and with the CCSD method in EOM-CCSD. Needless to say, the CCSD method includes more dynamic correlation than the (D) method but clearly the error cancellation between S$_1$ and S$_0$ is good unlike that for S$_0$ and T$_1$. 
Double excitations in the S$_0$ and S$_1$ states are thought to contribute about $~10\%$, according to the amplitudes of the CIS(D), EOM-CCSD and ADC(2) wave functions. Now, for the T$_1$ state, CIS(D) is not able to include sufficient dynamic correlation as we have discussed above.
The S$_1$ and T$_1$ states (of azine-1N, for example) have about $~5-7\%$ of doubly excited configurations with respect to S$_0$ (which are mostly singles out of their respective reference functions) if one were to allow a multireference description (see Fig.~\ref{weightage}). The dynamic correlation of these configurations cannot be accounted for in a single-reference SD framework but can be included with triples (or alternately, with doubles out of a multi-reference function). Hence, the triples in the $\Delta$CCSD(T) make-up for the lack of inclusion of other singly-excited configurations in the reference function.


The ADC manifold of excited state theories performs better in estimating the $\Delta$E$_{ST}$ than the individual excitation energies presumably due to error cancellations correct to a certain order of perturbation. In ADC(2) for instance, the ground state and all excited states are obtained correct up to second order of perturbation. The Hilbert space covered by ADC(2) and ADC(3) are same as CIS(D) but the balancing act in the coupling terms between the various excitation sub-spaces predicts more accurate excitation energies through error cancellation. ADC(3) shows an improvement in $\Delta$E$_{ST}$ with MP3 level ground and excited states. This behaviour is similar to that of CASSCF with smaller active spaces as can be seen in Table.~S3 from the supplementary material.

The $\Delta$CC methods apply a CCSD ansatz on the Hartree-Fock function of the chosen state. For the S$_1$ state this can be obtained by a maximum-overlap algorithm\cite{gilbert2008} when computing the reference Hartree-Fock state, for instance, while S$_0$ and T$_1$ are the variational minima of their respective spin multiplicities. 
The UHF numbers (of azine-1N, for example) from the supplementary material indicates that the energy of the MOM-UHF for S$_1$ is very close to S$_0$ but with an $<S^2>$ value of $~2.01$, while T$_1$ has $<S^2>=2.49$. Spin contamination of S$_1$ is thus, larger than T$_1$. $\Delta$HF underestimates the $S_1-S_0$ gap but is reasonable for the $T_1-S_0$ gap leading to a large negative $\Delta$E$_{ST}$~-0.98 eV. On the other hand, $\Delta$MP2 gives a large positive $\Delta$E$_{ST}$ due to overestimating both the $S_1-S_0$ and $T_1-S_0$ excitation energies. However orbital optimized (OO)-MP2 is able to predict the T$_1-$S$_0$ gap (~1.02 eV at OO-MP2/def2-TZVP) quite accurately recovering the spin of the T$_1$ state with a $<S^2>$ value of 2.01. Incidentally spin-opposite scaling (SOS) correction to OO-MP2 has a poorer outcome of 1.26 eV as expected. 

The $\Delta$CCSD method has the additional limitation of the absence of error cancellation (a strength of linear response methods) which makes it all the more crucial to include all of the correlation. UCCSD on the spin broken UHF references improves the $<S^2>$\cite{Stanton1994a} (from 2.29 (UHF/cc-pVDZ) for the T$_1$ to 2.014 (UCCSD/cc-pVDZ) in azine-1N reported in ref\cite{dreuw2023}) , but the $\Delta$E$_{ST}$ result is still very poor due to the spin contamination of S$_1$ ($<S^2>$ = 1.042). It is interesting to note that the UHF-CCSD/cc-pVDZ computation for the T$_1$ state of azine-1N which has the almost correct $<S^2>$ value of $2.014$ (ideally $<S^2> = 2$) also shows an accurate T$_1-$S$_0$ energy of 1.19 eV (EOM-CCSD/cc-pVDZ, $E_{T_1}$ =1.201 eV and ROHF-CCSD/cc-pVDZ, $E_{T_1}$ =1.19 eV). One may thus attribute the failure of the $\Delta$CCSD method to the spin-contamination of the reference determinant and the deficiencies of the reference function. Remarkably, in the UCCSD(T)\cite{Raghavachari1989,Stanton1997} the inclusion of the triples, (T), improves the energy differences greatly presumably due to the correction of the spin contamination along with the need to make up for the partial multireference nature of the reference function as discussed earlier. 

On the other hand, the ROSGM\cite{hait2016} formalism for HF, MP2 and CCSD uses a spin pure reference determinant resulting in excited SCF states that are orthogonal. We have seen that $\Delta$E$_{ST}$ values for ROSGM-HF are always positive (contrary to the MOM-UHF). But ROSGM-MP2 erroneously leads to a large negative ST gap. ROSGM-CCSD subsequently recovers values close to UHF-CCSD for the T$_1$ state possibly through introduction of sufficient correlation and some degree of spin contamination from the non-linear unrestricted ansatz. The S$_1$ excitation energies continue to be poor for ROSGM-CCSD leading to a large positive $\Delta$E$_{ST}$.

\begin{figure} 
\includegraphics[width=\textwidth]{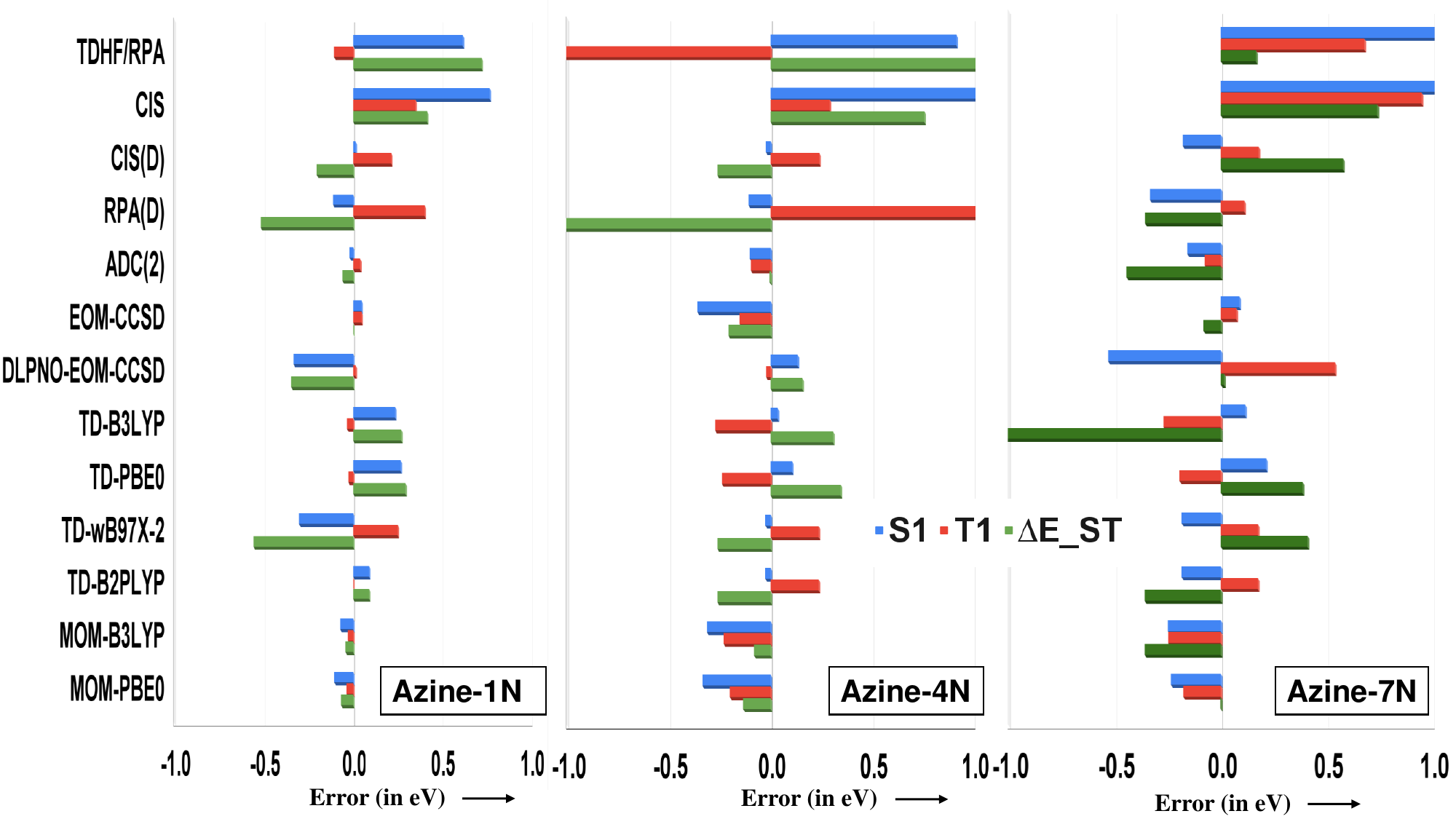}
\caption{Differences in the S$_1$-S$_0$, T$_1$-S$_0$ and $\Delta E_{ST}$ energy values in eV for Azine-1N, Azine-4N and Azine-7N computed with various single-reference correlation theories with FICMRCI (12,9) as the benchmark value}
\label{1N,4N,7N}
 \end{figure}

In Fig.~\ref{1N,4N,7N} we plot the excitation energies and $\Delta E_{ST}$ for azine-1N, azine-4N and azine-7N respectively computed with various single reference theories and also present some popular density-based methods for comparison and additional insight.
Interestingly, using the traditional LR-TDDFT technique with standard exchange-correlation functionals, positive singlet-triplet gaps ($\Delta E_{ST}>0$) are obtained but through a $\Delta$SCF approach such as the maximum overlap method (MOM)\cite{gilbert2008} a negative ST gap is obtained for the same XC functionals (eg. B3LYP and PBE0). This identifies the cause of the failure of LR-TDDFT for inverted singlet-triplet gaps as the absence of state-specific correlation coupled with orbital optimisation in the excited states which the MOM technique can capture. 
	By introducing spin-breaking in the KS determinant, we mimic part of the effect of spin-polarization\cite{drwal2023} but in an uncontrolled manner. We believe that the good performance of MOM-based DFT methods lies in introducing the spin-polarization through a coupling with the exchange-correlation functionals rather than the spin-breaking itself, similar to that observed in OO-MP2 for the T$_1$ state.
 
 In fact, removing the spin-contamination, through adopting, for example ROKS formalism\cite{hait2016} one is back to positive ST gaps\cite{ye2020arxiv} (eg. UB3LYP gives ST gap -0.12 eV with $<S^2>$ value 1.09 and 2.011 for S$_1$, T$_1$, while ROKS-B3LYP gives ST gap ~0.13 eV). The only difference here is the $<S^2>$ value (spin-contamination) clearly indicating its role in inversion of the ST gap during state specific orbital optimization. We must also note that MOM-DFT methods are much less spin-contaminated than UHF and KS orbitals have recently been adopted for post-SCF computations based on this argument\cite{rettig2020,fang2016,beran2003,MALLICK2021113326}. There are other strategies within DFT to include orbital orbital relaxation effects in the variational excited state solutions\cite{hait2016,hait2020,hait2021} which have a comparable performance vis a vis the MOM approach. Adding additional correlation through a post-KS-TDDFT state-specific MP2-like correction to the excited state energies in the so-called double hybrid XC functional based TD-DFT also captures the inversion\cite{ghosh2022,sanchogarchia2022,mojtaba2022}. Among the various double hybrid functionals $\omega$B97X-2 appears to perform poorly across all molecules. The others we have studied have similar and low errors. We try to understand this further along with the success of the Kohn-Sham $\Delta$SCF methods using a larger test set of molecules  
in a forthcoming partner publication\cite{saha2023}. We do not expect quantitative accuracy from KS-DFT but rather a consistent protocol for screening a large number of molecules.  

In this section we have seen that inclusion of dynamic correlation through single reference correlated wave function based methods - CIS(D), ADC(2) or EOM-CCSD, $\Delta$CCSD or $\Delta$SCF approaches through DFT, is sufficient to obtain inverted singlet-triplet ordering. However, the accurate estimation of the singlet-triplet near degeneracy seen in experiments is only achieved in EOM-CCSD and $\Delta$CCSD(T) indicating the need for either including static correlation or somehow balancing the amount of dynamic correlation in the states involved. Spin polarization through correlation is possibly the dominant physical effect that needs to be captured. The average $D_1$ diagnostics\cite{JANSSEN1998423} value for the S$_1$ state of the three systems at the CC2 level (as reported in\cite{ricci2021}), is 0.08, which are marginally greater than the $D_1 \leq 0.05$ criterion for single-reference character. The (GS) $T_1$-diagnostic\cite{lee1989} for azine-7N yields a value of 0.02 at CCSD/aug-cc-pVTZ level as reported in\cite{loos2023}, indicating absence of multi-reference (MR) character. Furthermore, $\%T1$ values of 86.3\% and 95.7\% are reported\cite{loos2023} for the lowest excited singlet and triplet states, respectively, from CC3/aug-cc-pVDZ computations. This indiactes dominance of singly excited character but the S$_1$ state has a significant contribution from doubly excited configurations. The lack of full orbital relaxation in the CCSD function (since the CC ansatz only modifies the ket function) can also lead to different conclusions of excitation character from CC-based and CASSCF based methods.
In any case, to further understand the relative weights of various configurations in these states we continue our investigations with a suite of multi-reference correlation theories in the next section.

\subsection{Multi-reference framework} \label{multiref}



\begin{figure}[h]
\centering
\begin{minipage}{\textwidth}
	\includegraphics[width=\textwidth]{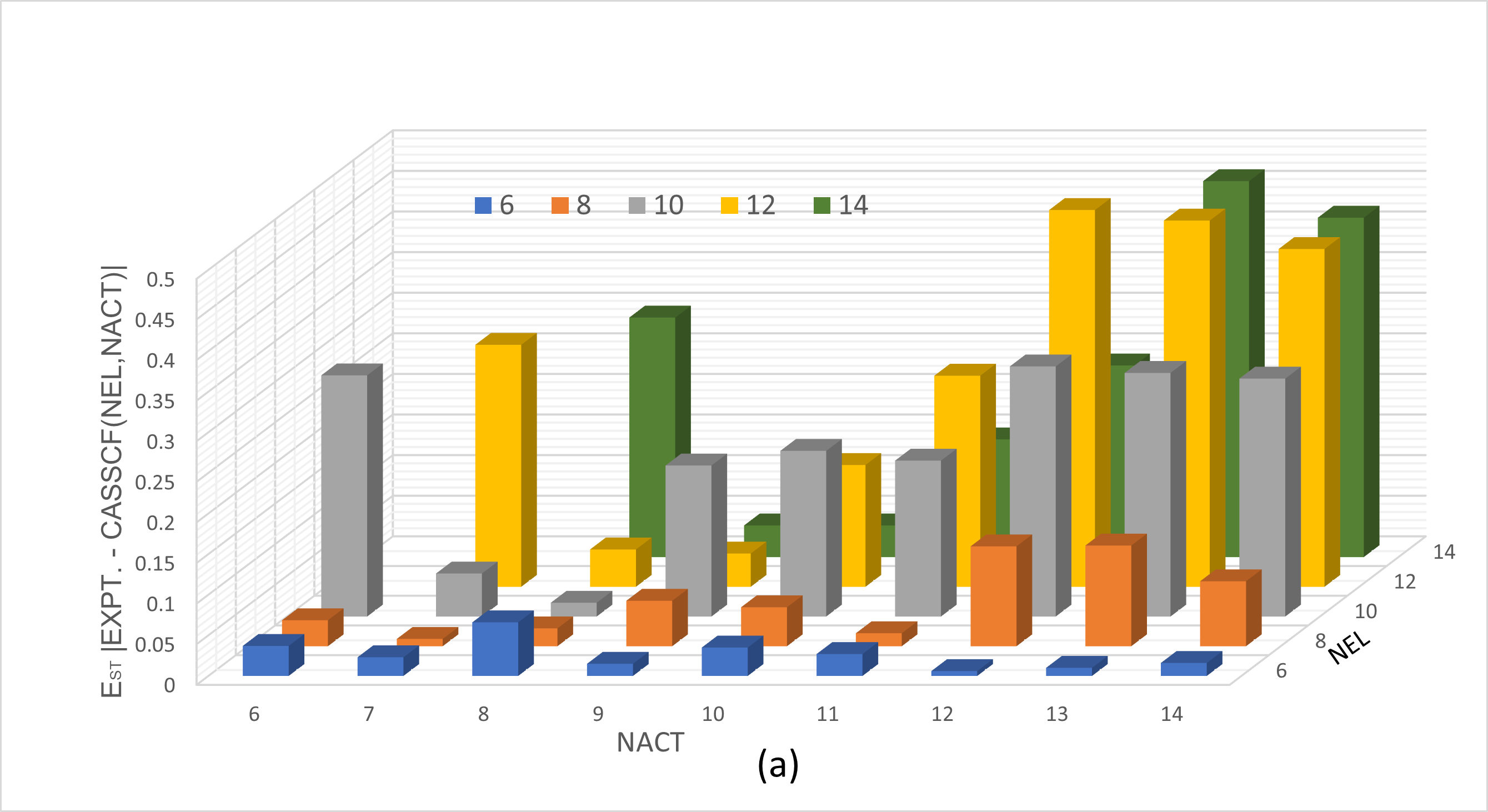}
 	\includegraphics[width=\textwidth]{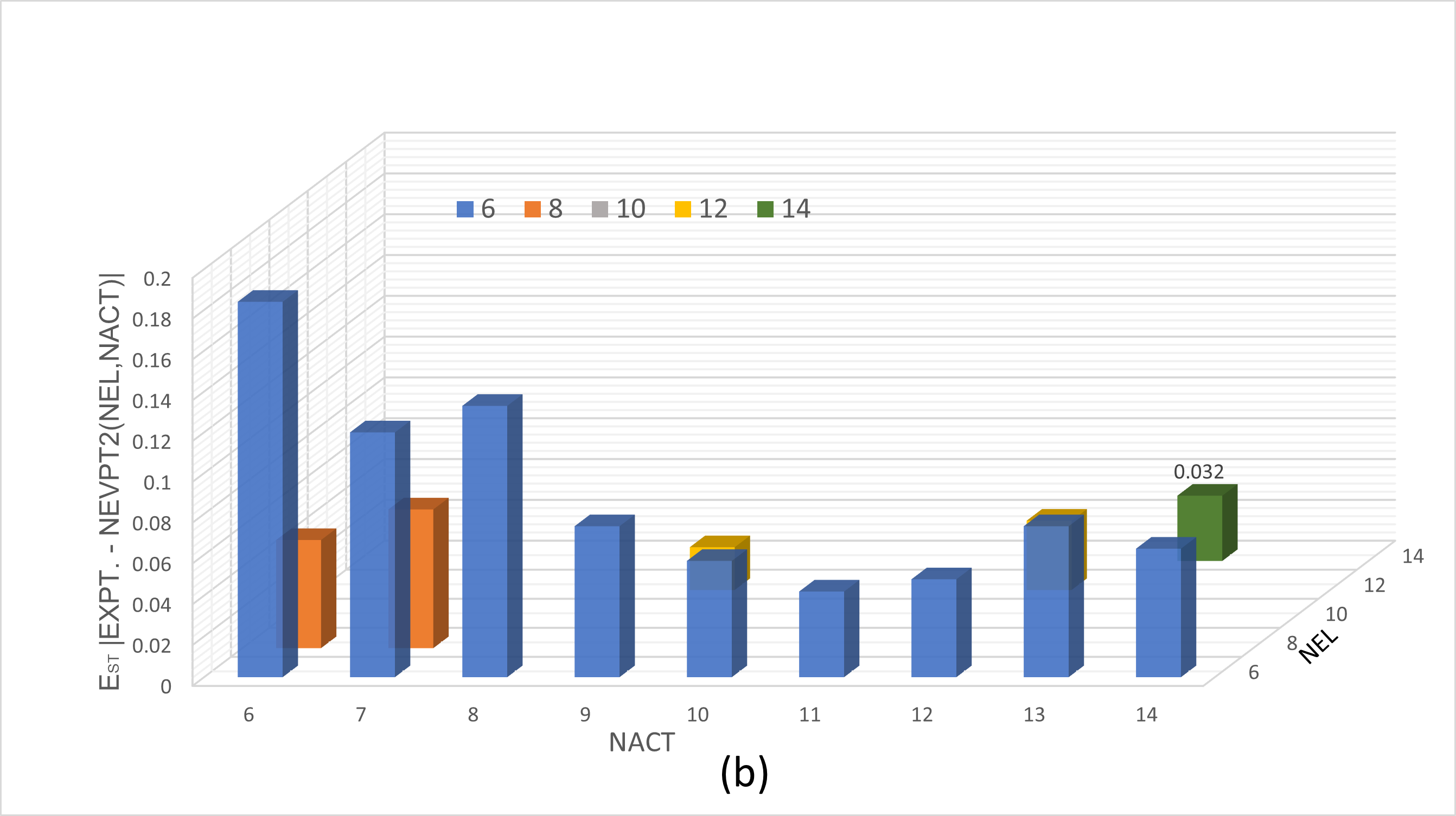}
\end{minipage}
\caption{Deviation of CASSCF results with experimental value (in eV) for varying active space and NEVPT2 values for selected active space for azine-1N}
\label{CAS}
\end{figure}

For the multi-reference treatment of any electronic state, a proper choice of `active space' is necessary. As discussed in Sec.~\ref{theory}, this involves choosing a set of `active orbitals' from a previous single-reference SCF computation. The active orbitals need to be chosen through experience and intuition as no definite rule is available. Guidelines and automated tools for active space selection are a field of ongoing research\cite{bao2019,golub2021,stein2019}. There are roughly-speaking two common practices - energy based and phenomenon based - although several new strategies are recently being investigated\cite{king2021,veryazov2021}.
In the energy-based strategy one selects quasi-degenerate orbitals from among the occupied and unoccupied molecular orbitals (MOs). This becomes progressively difficult as the size of the molecule increases and a clear cut-off cannot be obtained in the orbital energies. Moreover, the success of this strategy depends very much on the quality of the MOs in the starting calculation. In the phenomenon-based strategy one may choose the orbitals involved in a bond-breaking process as the active orbitals for constructing a potential energy curve or, as in our case, the class of orbitals likely to be involved in a particular type of excitation. Since the Hartree-Fock reference function is qualitatively wrong in our case, we have carried out a thorough study of the possible active spaces before arriving at the correct choice for further correlation studies. We discuss this below.

In Table~S3 in the supplementary material we report the excitation energies and ST gaps of azine-1N computed using CASSCF with various choices of active electron ($N_{el}$) and active orbitals ($N_{act}$) chosen in energy ordering (with a plot of the values in the Fig~S2). The errors in the $\Delta$E$_{ST}$ against the experimental value of $-0.07$ eV of azine-1N along with NEVPT2 error plot for selected active spaces are presented in the segment (a) and (b) of Fig.~\ref{CAS} respectively. It is evident from the values, that a naive increase of active orbitals and electrons does not help in finding the correct active space. The lack of convergence of the CASSCF excitation energies with increasing active space can be an indication of poor starting orbitals and/or significant dynamic correlation\cite{king2021}. 
From a phenomenon-based point of view, all $\pi$ and $\pi^*$ orbitals from the RHF calculation of azine-1N should be considered as active orbitals along with the lone pair orbital of N. For azine-1N there are 6 $\pi$ bonds and 1 lone pair orbitals which are occupied and 6 $\pi^*$ and 1 or more N p$_z$ containing unoccupied orbitals. The 6 $\pi$ orbitals from HOMO-5 to HOMO and 3 $\pi^*$ valence orbitals, LUMO to LUMO+2, are energetically in order and can be included in a CAS(12,9) (12 electrons, 9 orbitals) calculation. The rest are reshuffled in the RHF calculation. Beyond the (12,9) space, we have identified the N lone pair and reshuffled the orbitals to form the (14,10) CAS. Subsequently, we identified all the likely unoccupied active orbitals and did a systematic increase of active space for only one case (azine-1N) upto a CAS(14,14) which is the largest active space we have considered. These results are presented in Table~S5 in the supplementary material. In the same table, we also present the excitation energies after inclusion of dynamic correlation on these active spaces.
This is important to compare because for the smaller active spaces such as (6,n) or (8,n), etc which do not have any logical basis for being a good choice, there is an apparent accuracy in the energy values at the CASSCF level (Fig.~\ref{CAS}(a)). But, further inclusion of correlation distorts the picture indicating that the agreement was fortuitous as borne out by the NEVPT2 error plot in the Fig.~\ref{CAS}(b). The inclusion of the N lone pair orbital led to marginal improvement in azine-1N. Inclusion of the remaining $\pi^*$s which are energetically even higher than some $\sigma*$ orbitals led to a deterioration of the results presumably due to the poor quality of the HF MOs. To cross-check our reasoning, we have carried out CASSCF (14,10) computations using a reference determinant constructed using CASSCF (14,14) orbitals (see Table~S5 of the supplementary material). There is a marked improvement in this case taking the excitation energies closer to the FIC-MRCI (14,10) values. We may thus assert that a significant portion of the role of post-CASSCF corrections comes from state-specific orbital relaxation. This is also an indirect argument for the success of $\Delta$SCF DFT methods using MOM/SGM techniques (see Fig~(\ref{1N,4N,7N})). In summary, CAS(12,9) where 6 occupied $\pi$ and 3 virtual $\pi^*$ are included appears to be a good choice as it satisfies both the energetic criteria and the phenomenon-based criteria in the active space. In the case of the other systems, the RHF computation gives rise to 3 occupied $\pi$ and 3 virtual $\pi^*$ orbitals ordered energetically while the other $\pi$ and $\pi^*$ orbitals are reshuffled. Including other $\pi$ orbitals through swapping we have extended the active space to (12,9) and then to (14,10) by including N-lone pair. Pictures of the active orbitals taken for the CAS(12,9) and CAS(14,10) CASSCF calculations are presented in Fig.~S1 of the supplementary material along with the natural occupation numbers of the fractionally occupied active orbitals for state-specific CASSCF(12,9) calculations for S$_0$, S$_1$ and T$_1$ states in Table~S2. Errors in all the multi-reference theories against the experimental values for azine-1N with different active spaces are plotted in Fig.~S5 in the supplementary material.

\begin{figure}[h]
\centering
\begin{minipage}{\textwidth}
	\includegraphics[width=0.45\textwidth]{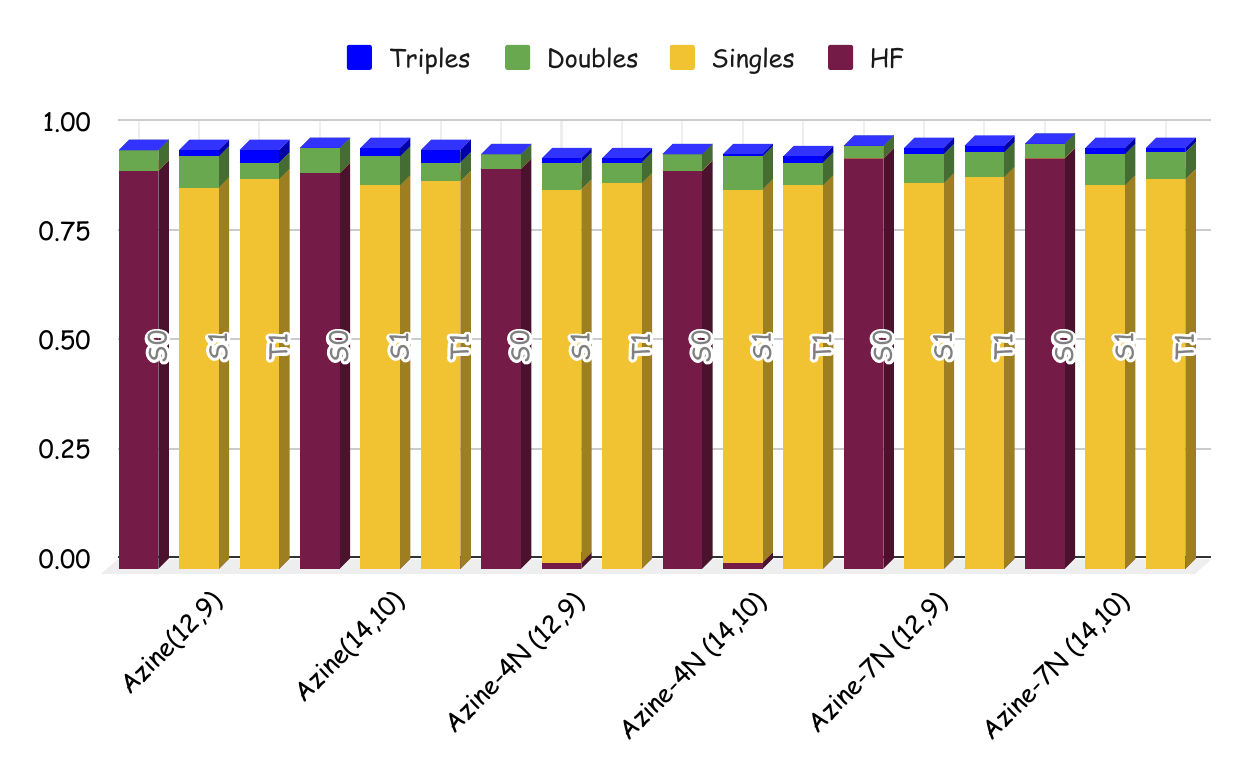}
 	\includegraphics[width=0.45\textwidth]{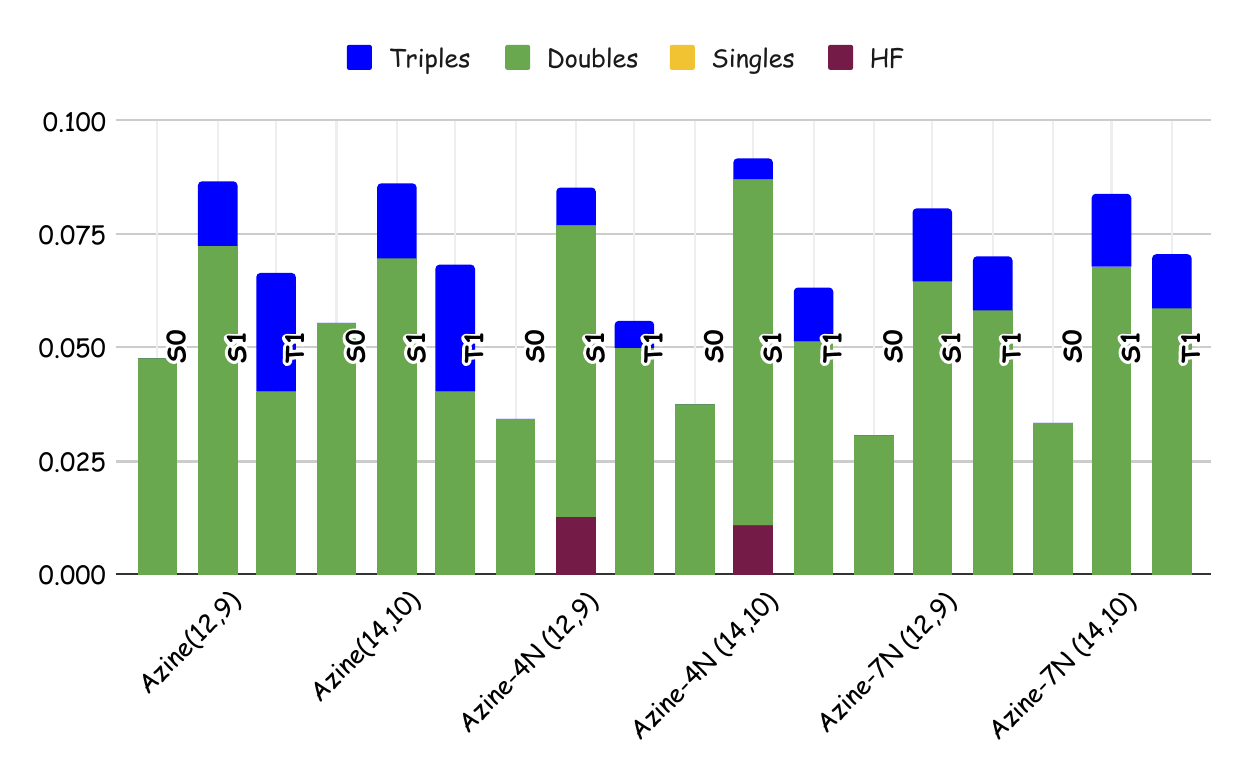}
\end{minipage}
\caption{Fraction of n-tuple excitaions in CASSCF wavefunction of Azine-1N, Azine-4N and Azine-7N. The right-panel highlights the non-singles contribution to each state.}
\label{weightage}
\end{figure}

 To further understand the nature of the static correlation, we have studied the weightage of double- and higher-order excitations in the final CAS wavefunction of these systems. However, while evaluating the double excitation character of the S$_1$ state, the multi-reference nature of the ground state must be taken into consideration. Thus we refer to the RHF function of the S$_0$ state as the reference determinant for classifying singles, doubles, triples, etc. In the following discussion we generally refer to the computation with a (12,9) CAS but the corresponding values for the (14,10) CAS are also presented in Fig.~\ref{weightage}. For all the molecules, the HF configuration makes up about 90\% of the ground-state CASSCF wave function but it also contains a significant contribution (5\%) from the doubly excited configurations. With singly excited configurations contributing between 80\% - 87\%, the S$_1$ and T$_1$ excited state is primarily singly excited in nature with the largest contribution coming from a single CSF. The multi-reference nature of S$_1$ and T$_1$ comes from the doubles which contribute $~5-7\%$ and triples which contribute $1-2\%$.  
 On comparing the CASSCF configurations of the (12,9) and (14,10) active spaces, no significant new configurations are seen. However, the T$_1-$S$_0$ excitation energy reduces by $0.012$ eV indicating partial inclusion of dynamic correlation rather than static correlation. The weightage of the singles, doubles and triples contributions in the CASSCF wave functions for the two different CAS spaces are pictorially given in Fig.~\ref{weightage}.
If we consider the importance of n-tuple excitons beyond single-particle ones, it should be noted that values of ~5\%, ~3\%, and ~3\% for azine-1N, azine-4N, and azine-7N, respectively, are estimated for the S$_0$ wavefunction. For the corresponding S$_1$ (T$_1$) excited-states, these weights of the doubles are about $~6-7\%$ ($~5-6\%$) respectively. Given that Coulomb correlation predominantly involves the interaction between anti-parallel spins, it is known that doubly-excited configurations result in stabilisations that are a few times bigger for singlet than for triplet excited-states. This leads to greater relative lowering of the S$_1-$S$_0$ gap vis a vis the T$_1-$S$_0$ gap. 
In short, ground state of these molecules are predominantly single-reference (as evident from the (GS) $T_1$ diagnostic value\cite{loos2023}), S$_1$ and T$_1$ have moderate multi-reference character. 

We have included dynamic correlation on the CASSCF wave-function, along with the PT2 methods - SC-NEVPT2 and FIC-CASPT2 and also the non-perturbative FIC-MRCI. 
SC-NEVPT2 predicts that the singlet-triplet gaps of all molecules would either be negative or very near to zero similar to EOM-CCSD. In all cases FIC-CASPT2 underestimates the energy gap. As shown in Table.~S5 in the supplementary material, the CASSCF values are very poor for large active spaces even for (14,14) where all $\pi-\pi^*$ orbitals are considered. It indicates that the introduction of non-dynamical correlation effects is only a first step in dealing with the complex electronic effects of triangle-shaped systems. We can also see in Table~S5, that NEVPT2 can recover a reasonable $\Delta$E$_{ST}$ from extremely poor CAS(12,12) and CAS(14,13) CASSCF values. This supports our suggestion that a static+dynamic correlation approach is less sensitive to the choice of the active space. Incidentally CASPT2 improves with increase in size of the active space.  
SC-NEVPT2, FIC-CASPT2 and FIC-MRCI results are given in Table~\ref{MRcorr} for two different active spaces - CAS (12,9) and (14,10). 
The SA-CASSCF results for the S$_1$ and T$_1$ excitation energies of azine-7N are significantly reduced by the SC-NEVPT2 correction, yielding remarkably consistent results in all situations.  
Also take note of the quantitative agreement between SC-NEVPT2 values and experimental estimates for specific S$_1$ and T$_1$ excitation energies (which have been previously reported by benchmark calculations of tiny systems against pseudo-FCI\cite{loos2019}). 
It can be noted that for all cases, FIC-CASPT2 values are poor for selected active spaces and sometimes underestimates the exciton energies. But it is evident from the Table~S5 of the supplementary material that, inclusion of more $\pi^*$ orbitals beyond (14,10) CAS, leads to error in NEVPT2, but improves CASPT2 values. Infact, in the case of CAS(14,14), the FIC-CASPT2 value is very accurate and close to the experimental value. It is also reflected in the plot of the values presented in Fig.~S5.

\begingroup                      
\begin{table}
\begin{ruledtabular}
\caption{\label{MRcorr} Vertical excitation energies (in eV) for CAS(12,9) and CAS(14,10) with SC-NEVPT2, FIC-CASPT2 and FIC-MRCI. Geometries are optimized at B3LYP/def2-TZVP level.}
\begin{tabular}{ c  c   c   c  c  c  c  c}
& & &CAS(12,9) & & & CAS(14,10) & \\
\hline
Molecule    & Method & S$_1$ & T$_1$ & $\Delta$E$_{ST}$ & S$_1$ & T$_1$ & $\Delta$E$_{ST}$ \\
\hline
  Azine-1N  
&CASSCF & 0.927 &	0.956 &	-0.029 & 0.974 & 1.005 & -0.031 \\
&NEVPT2 & 0.964 & 1.055 & -0.091	& 0.939 & 1.043 & -0.104 \\
&CASPT2 & 0.413 & 0.656 & -0.243 & 0.357	& 0.618 & -0.261 \\
&FICMRCI & 1.022 & 1.099	& -0.077 & 1.034 & 1.118 & -0.084 \\
   &Expt$^b$ & 0.97 & 0.93/1.05 & 0.04/-0.07 & 0.97 & 0.93/1.05 & 0.04/-0.07 \\
\hline
 Azine-4N
&CASSCF & 2.212 & 2.191 & 0.021 & 2.288 & 2.276 & 0.012 \\
&NEVPT2 & 2.246 & 2.231 & 0.015 & 2.27 &	2.256 &	0.014 \\
&CASPT2 & 1.585 & 1.71 &	-0.125	& 1.611 & 1.72 & -0.109 \\
&FICMRCI & 2.249	& 2.235	& 0.014	& 2.301 & 2.294 & 0.007  \\
   &Expt & $<$2.39 & 2.29 & $<$0.1 & $<$2.39 & 2.29 & $<$0.1\\
\hline
 Azine-7N 
&CASSCF & 2.841 & 2.999 & -0.158 & 2.992 & 3.124 & -0.132 \\
&NEVPT2	& 2.59	& 2.77	& -0.18	& 2.504 &	2.776 &	-0.272 \\
&CASPT2	& 1.812	& 2.234	& -0.422 & 1.715 &	2.202 &	-0.487 \\
&FICMRCI	& 2.829	& 2.994	& -0.165 & 2.9 & 3.097 & -0.197 \\
   &Expt & $--$ & $--$ & $<$0 & $--$ & $--$ & $<$0\\
\end{tabular}
\end{ruledtabular}
\end{table}
\endgroup


The FIC-MRCI includes a greater degree of dynamic correlation. From the FIC-MRCI results given in Table~II for CAS (12,9) and (14,10), we can conclude that the S$_1$ and T$_1$ excitons significantly differ sometimes from the SC-NEVPT2 values. The dynamic correlation contributions to the excited state energies recovered through FIC-MRCI framework are about 10.24\%, 1.9\% and 6\% in S$_1$ state for azine-1N, azine-4N and azine-7N respectively and 15\%, 1\% and 4.9\% for the T$_1$ states. The percentage contribution of dynamic correlations captured by both the frameworks are similar for azine-1N, but deviate a lot for azine-4N and azine-7N. This indicates that these two molecules have large dynamical correlations.  

We may thus opine that a benchmark method for this class of molecules should involve both static and dynamic correlation such as SC-NEVPT2 or FIC-MRCI such that the description of correlation is complete and hence, reliable across different molecules alongwith the advantage of less sensitivity to the choice of the active space (compared to CASSCF). EOM-CCSD is also quite good for the molecules studied here but from a theoretical perspective it misses the differential correlation of ground and excited states and may lead to reduced accuracy in other molecules.




\section{Summary and Future Outlook} \label{summary}
The lowest singlet S$_1$ and triplet T$_1$ excited states of a group of N-doped (triangular-shaped) -conjugated hydrocarbons, namely azine-1N, azine-4N, and azine-7N, are examined for their energy ordering. In these systems the S$_1-$T$_1$ gaps are predicted to be negative, which will encourage exoergic reverse intersystem crossing events, which are useful in various optical device technologies such as OLEDs.
Regardless of the exchange-correlation functional selected, the S$_1-$T$_1$ gap is always found to be positive when using the TD-DFT approach with standard exchange-correlation functionals. However, with the logical exception of CIS, every correlated wave-function approach investigated, such as CIS(D), EOM-CCSD, ADC(2), CASSCF, SC-NEVPT2, and FIC-MRCI resulted in $\Delta E_{ST} \leq 0$ values for these systems. 


While most previous studies have focused on which methods predict a larger inversion, we have found that higher levels of theory actually predict near degeneracy of S$_1$ and T$_1$ with $ -0.2 < \Delta E_{ST} < 0.01$ eV which is closer to experimental values where available. This is heartening because a large negative energy gap can also be detrimental to an efficient reverse intersystem crossing\cite{dinkelbach2021}. 
We have identified the delicate balance between static and dynamic correlation which takes us from the large positive $\Delta E_{ST}$ of uncorrelated wavefunction theories to the large negative $\Delta E_{ST}$ of CIS(D) and RPA(D) to the near degeneracy of EOM-CCSD, CASSCF, NEVPT2 and FIC-MRCI.
A careful consideration of the CASSCF coefficients have indicated that while S$_0$ is single-reference, the S$_1$ and T$_1$ states are moderately multi-reference with about 5-15\% contributions from configurations other than the reference for the respective states of the three molecules. The effect of these additional configurations can thus, be captured with a large amount of dynamical correlation or with a static plus dynamic approach. We have observed that the correlation of the T$_1$ state is only slightly weaker (not dramatic as surmised by others) than that of the S$_1$ which brings the state closer with a small inverted or near-zero gap. This difference is likely to arise from spin-dependent orbital relaxation and spin-dependent correlation relaxation\cite{drwal2023}. Theories which capture the latter effects properly are found to give the most accurate results. In EOM-CCSD, the so-called R operators introduce spin-polarization of reference orbitals in the presence of correlation (from the T operators) within the single-reference framework. This coupling is absent in CIS(D) and RPA(D), for instance, which only include correlation additively. In CASSCF-based multireference approaches, on the other hand, correlation is included individually on spin-adapted references leading to automatic inclusion of these effects. One must note that while smaller active spaces in CASSCF sometimes give accurate numbers, subsequent inclusion of dynamic correlation may distort the trends leading to our suggestion of the (12,9) active space as being numerically accurate as well as being physically meaningful. 

We have also explored the possibility of computing the singlet-triplet energies of our test set using the $\Delta$SCF and $\Delta$CCSD class of methods in order to observe the impact of state-specific orbital optimisation on the excited-state energetics. Our findings indicate that spin-contamination of the open-shell states play a very important role and can completely overshadow any gains from inclusion of state-dependent orbital optimization and correlation. This class of molecules show a very high spin contamination at the UHF level even for the T$_1$ state and subsequent inclusion of dynamic correlation through CCSD improves it but accurate numbers can only be achieved at the CCSD(T)\cite{Raghavachari1989} level. On the other hand, UDFT methods show very little (if any) spin contamination for the variational minima states but significant spin-contamination is seen for the artificially converged higher excited states. However, this is still lower than that in the corresponding UHF excited state. Preliminary investigations indicate that $\Delta$SCF DFT methods do very well for these molecules and we shall follow up with a more extensive benchmark study of DFT-based methods. These findings point to the necessity of spin-dependent correlation-sensitive orbital optimisation, which the LR-TDDFT lacks, in order to achieve accurate singlet-triplet gaps in these compounds. We have also found that the TDDFT(D) approach and the appropriate double-hybrid density functionals perform well.

Thus, this paper identifies the nature of the electron correlations which affect the lowest excitation energies of molecular templates for potential OLED materials and tries to explain the success and failure of various wavefunction-based ab initio electronic structure theories for computing inverted or nearly degenerate S$_1$ and T$_1$ excited states. EOM-CCSD, FIC-NEVPT2 and FIC-MRCISD have been found to be suitable benchmark methods. The density-functional approaches are to be discussed and benchmarked in a follow-up study such that one or more accurate yet computationally cheap theories can be identified for screening relatives and derivatives of the templates studied here.

\section*{Supporting Information}
State energies of S$_0$, S$_1$ and T$_1$ for azine-xN along with the cartesian geometries and pictures of the active orbitals used are recorded. Data from additional studies using multi-reference methods discussed in the paper are presented along with some plots to support the arguments in the paper.

\section*{Acknowledgements}
SS gratefully acknowledges funding support from DST SERB, New Delhi, India (SRG/2021/000737) and IISER Kolkata. SC want to thank IISER Kolkata for Junior Research Fellowship. Prof. Ashwani K. Tiwari is acknowledged for sharing his computational facilities. We thank the two anonymous referees for their valuable comments.  
\bibliographystyle{achemso}
\bibliography{invest.bib}

\providecommand{\latin}[1]{#1}
\makeatletter
\providecommand{\doi}
  {\begingroup\let\do\@makeother\dospecials
  \catcode`\{=1 \catcode`\}=2 \doi@aux}
\providecommand{\doi@aux}[1]{\endgroup\texttt{#1}}
\makeatother
\providecommand*\mcitethebibliography{\thebibliography}
\csname @ifundefined\endcsname{endmcitethebibliography}
  {\let\endmcitethebibliography\endthebibliography}{}
\begin{mcitethebibliography}{98}
\providecommand*\natexlab[1]{#1}
\providecommand*\mciteSetBstSublistMode[1]{}
\providecommand*\mciteSetBstMaxWidthForm[2]{}
\providecommand*\mciteBstWouldAddEndPuncttrue
  {\def\EndOfBibitem{\unskip.}}
\providecommand*\mciteBstWouldAddEndPunctfalse
  {\let\EndOfBibitem\relax}
\providecommand*\mciteSetBstMidEndSepPunct[3]{}
\providecommand*\mciteSetBstSublistLabelBeginEnd[3]{}
\providecommand*\EndOfBibitem{}
\mciteSetBstSublistMode{f}
\mciteSetBstMaxWidthForm{subitem}{(\alph{mcitesubitemcount})}
\mciteSetBstSublistLabelBeginEnd
  {\mcitemaxwidthsubitemform\space}
  {\relax}
  {\relax}

\bibitem[Leupin and Wirz(1980)Leupin, and Wirz]{jacs80}
Leupin,~W.; Wirz,~J. Low-Lying Electronically Excited States of
  Cycl[3.3.3]azine, a Bridged 12$\pi$-Perimeter. \emph{J. Am. Chem. Soc.}
  \textbf{1980}, \emph{102}, 6068\relax
\mciteBstWouldAddEndPuncttrue
\mciteSetBstMidEndSepPunct{\mcitedefaultmidpunct}
{\mcitedefaultendpunct}{\mcitedefaultseppunct}\relax
\EndOfBibitem
\bibitem[Leupin \latin{et~al.}(1986)Leupin, Magde, Persy, and Wirz]{jacs86}
Leupin,~W.; Magde,~D.; Persy,~G.; Wirz,~J. 1, 4, 7-Triazacycl [3.3. 3] azine:
  basicity, photoelectron spectrum, photophysical properties. \emph{J. Am.
  Chem. Soc.} \textbf{1986}, \emph{108}, 17\relax
\mciteBstWouldAddEndPuncttrue
\mciteSetBstMidEndSepPunct{\mcitedefaultmidpunct}
{\mcitedefaultendpunct}{\mcitedefaultseppunct}\relax
\EndOfBibitem
\bibitem[Endo \latin{et~al.}(2011)Endo, Sato, Yoshimura, Kai, Kawada, Miyazaki,
  and Adachi]{endo2011}
Endo,~A.; Sato,~K.; Yoshimura,~K.; Kai,~T.; Kawada,~A.; Miyazaki,~H.;
  Adachi,~C. Efficient up-conversion of triplet excitons into a singlet state
  and its application for organic light emitting diodes. \emph{Appl. Phys.
  Lett.} \textbf{2011}, \emph{98}, 083302\relax
\mciteBstWouldAddEndPuncttrue
\mciteSetBstMidEndSepPunct{\mcitedefaultmidpunct}
{\mcitedefaultendpunct}{\mcitedefaultseppunct}\relax
\EndOfBibitem
\bibitem[Uoyama \latin{et~al.}(2012)Uoyama, Goushi, Shizu, Nomura, and
  Adachi]{uoyama2012}
Uoyama,~H.; Goushi,~K.; Shizu,~K.; Nomura,~H.; Adachi,~C. Highly efficient
  organic light-emitting diodes from delayed fluorescence. \emph{Nature}
  \textbf{2012}, \emph{492}, 234\relax
\mciteBstWouldAddEndPuncttrue
\mciteSetBstMidEndSepPunct{\mcitedefaultmidpunct}
{\mcitedefaultendpunct}{\mcitedefaultseppunct}\relax
\EndOfBibitem
\bibitem[de~Silva \latin{et~al.}(2019)de~Silva, Kim, Zhu, and
  Van~Voorhis]{desilva2019a}
de~Silva,~P.; Kim,~C.~A.; Zhu,~T.; Van~Voorhis,~T. Extracting design principles
  for efficient thermally activated delayed fluorescence (TADF) from a simple
  four-state model. \emph{Chem. Mater.} \textbf{2019}, \emph{31}, 6995\relax
\mciteBstWouldAddEndPuncttrue
\mciteSetBstMidEndSepPunct{\mcitedefaultmidpunct}
{\mcitedefaultendpunct}{\mcitedefaultseppunct}\relax
\EndOfBibitem
\bibitem[Difley \latin{et~al.}(2008)Difley, Beljonne, and
  Van~Voorhis]{difley2008}
Difley,~S.; Beljonne,~D.; Van~Voorhis,~T. On the singlet-triplet splitting of
  geminate electron-hole pairs in organic semiconductors. \emph{J. Am. Chem.
  Soc.} \textbf{2008}, \emph{130}, 3420\relax
\mciteBstWouldAddEndPuncttrue
\mciteSetBstMidEndSepPunct{\mcitedefaultmidpunct}
{\mcitedefaultendpunct}{\mcitedefaultseppunct}\relax
\EndOfBibitem
\bibitem[Sato \latin{et~al.}(2015)Sato, Uejima, Tanaka, Kaji, and
  Adachi]{sato2015}
Sato,~T.; Uejima,~M.; Tanaka,~K.; Kaji,~H.; Adachi,~C. A light-emitting
  mechanism for organic light-emitting diodes: Molecular design for inverted
  singlet-triplet structure and symmetry-controlled thermally activated delayed
  fluorescence. \emph{J. Mater. Chem. C} \textbf{2015}, \emph{3}, 870\relax
\mciteBstWouldAddEndPuncttrue
\mciteSetBstMidEndSepPunct{\mcitedefaultmidpunct}
{\mcitedefaultendpunct}{\mcitedefaultseppunct}\relax
\EndOfBibitem
\bibitem[Di \latin{et~al.}(2017)Di, Romanov, Yang, Richter, Rivett, Jones,
  Thomas, Abdi~Jalebi, Friend, and Linnolahti]{di2017}
Di,~D.; Romanov,~A.~S.; Yang,~L.; Richter,~J.~M.; Rivett,~J. P.~H.; Jones,~S.;
  Thomas,~T.~H.; Abdi~Jalebi,~M.; Friend,~R.~H.; Linnolahti,~M.
  High-performance light-emitting diodes based on carbene-metal-amides.
  \emph{Science} \textbf{2017}, \emph{356}, 159\relax
\mciteBstWouldAddEndPuncttrue
\mciteSetBstMidEndSepPunct{\mcitedefaultmidpunct}
{\mcitedefaultendpunct}{\mcitedefaultseppunct}\relax
\EndOfBibitem
\bibitem[Olivier \latin{et~al.}(2018)Olivier, Sancho-García, Muccioli,
  D’Avino, and Beljonne]{olivier2018}
Olivier,~Y.; Sancho-García,~J.~C.; Muccioli,~L.; D’Avino,~G.; Beljonne,~D.
  Computational Design of Thermally Activated Delayed Fluorescence Materials:
  The Challenges Ahead. \emph{J. Phys. Chem. Lett.} \textbf{2018}, \emph{9},
  6149\relax
\mciteBstWouldAddEndPuncttrue
\mciteSetBstMidEndSepPunct{\mcitedefaultmidpunct}
{\mcitedefaultendpunct}{\mcitedefaultseppunct}\relax
\EndOfBibitem
\bibitem[Dinkelbach \latin{et~al.}(2021)Dinkelbach, Bracker, Kleinschmidt, and
  Marian]{dinkelbach2021}
Dinkelbach,~F.; Bracker,~M.; Kleinschmidt,~M.; Marian,~C.~M. Large Inverted
  Singlet–Triplet Energy Gaps Are Not Always Favorable for Triplet
  Harvesting: Vibronic Coupling Drives the (Reverse) Intersystem Crossing in
  Heptazine Derivatives. \emph{J. Phys. Chem. A} \textbf{2021}, \emph{125},
  10044\relax
\mciteBstWouldAddEndPuncttrue
\mciteSetBstMidEndSepPunct{\mcitedefaultmidpunct}
{\mcitedefaultendpunct}{\mcitedefaultseppunct}\relax
\EndOfBibitem
\bibitem[Aizawa \latin{et~al.}(2022)Aizawa, Pu, Harabuchi, Nihonyanagi, Ibuka,
  Inuzuka, Dhara, Koyama, Nakayama, Maeda, Araoka, and Miyajima]{nature2022}
Aizawa,~N.; Pu,~Y.-J.; Harabuchi,~Y.; Nihonyanagi,~A.; Ibuka,~R.; Inuzuka,~H.;
  Dhara,~B.; Koyama,~Y.; Nakayama,~K.-i.; Maeda,~S.; Araoka,~F.; Miyajima,~D.
  Delayed fluorescence from inverted singlet and triplet excited states.
  \emph{Nature} \textbf{2022}, \emph{609}, 502--506\relax
\mciteBstWouldAddEndPuncttrue
\mciteSetBstMidEndSepPunct{\mcitedefaultmidpunct}
{\mcitedefaultendpunct}{\mcitedefaultseppunct}\relax
\EndOfBibitem
\bibitem[Chanda \latin{et~al.}(2024)Chanda, Saha, and Sen]{saha2023}
Chanda,~S.; Saha,~S.; Sen,~S. Benchmark Computations of Nearly Degenerate
  Singlet and Triplet states of N-heterocyclic Chromophores: II. Density-based
  Methods. \emph{To be submitted} \textbf{2024}, \relax
\mciteBstWouldAddEndPunctfalse
\mciteSetBstMidEndSepPunct{\mcitedefaultmidpunct}
{}{\mcitedefaultseppunct}\relax
\EndOfBibitem
\bibitem[Hund(1925)]{hund1925}
Hund,~F. Zur deutung verwickelter spektren, insbesondere der elemente scandium
  bis nickel. \emph{Physik} \textbf{1925}, \emph{33}, 345\relax
\mciteBstWouldAddEndPuncttrue
\mciteSetBstMidEndSepPunct{\mcitedefaultmidpunct}
{\mcitedefaultendpunct}{\mcitedefaultseppunct}\relax
\EndOfBibitem
\bibitem[Bene \latin{et~al.}(1971)Bene, Ditchfield, and Pople]{pople2003}
Bene,~J. E.~D.; Ditchfield,~R.; Pople,~J.~A. {Self‐Consistent Molecular
  Orbital Methods. X. Molecular Orbital Studies of Excited States with Minimal
  and Extended Basis Sets}. \emph{J. Chem. Phys.} \textbf{1971}, \emph{55},
  2236--2241\relax
\mciteBstWouldAddEndPuncttrue
\mciteSetBstMidEndSepPunct{\mcitedefaultmidpunct}
{\mcitedefaultendpunct}{\mcitedefaultseppunct}\relax
\EndOfBibitem
\bibitem[Foresman \latin{et~al.}(1992)Foresman, Head-Gordon, Pople, and
  Frisch]{foresman1992}
Foresman,~J.~B.; Head-Gordon,~M.; Pople,~J.~A.; Frisch,~M.~J. Toward a
  systematic molecular orbital theory for excited states. \emph{J Phys. Chem.}
  \textbf{1992}, \emph{96}, 135--149\relax
\mciteBstWouldAddEndPuncttrue
\mciteSetBstMidEndSepPunct{\mcitedefaultmidpunct}
{\mcitedefaultendpunct}{\mcitedefaultseppunct}\relax
\EndOfBibitem
\bibitem[Bouman \latin{et~al.}(1983)Bouman, Hansen, Voigt, and
  Rettrup]{bouman1983}
Bouman,~T.~D.; Hansen,~A.~E.; Voigt,~B.; Rettrup,~S. Large-scale RPA
  calculations of chiroptical properties of organic molecules: Program RPAC.
  \emph{Int. J Quant. Chem.} \textbf{1983}, \emph{23}, 595--611\relax
\mciteBstWouldAddEndPuncttrue
\mciteSetBstMidEndSepPunct{\mcitedefaultmidpunct}
{\mcitedefaultendpunct}{\mcitedefaultseppunct}\relax
\EndOfBibitem
\bibitem[Bouman and Hansen(1989)Bouman, and Hansen]{bouman1989}
Bouman,~T.~D.; Hansen,~A.~E. Linear response calculations of molecular optical
  and magnetic properties using program RPAC: NMR shielding tensors of pyridine
  and n-azines. \emph{Int. J Quant. Chem.} \textbf{1989}, \emph{36},
  381--396\relax
\mciteBstWouldAddEndPuncttrue
\mciteSetBstMidEndSepPunct{\mcitedefaultmidpunct}
{\mcitedefaultendpunct}{\mcitedefaultseppunct}\relax
\EndOfBibitem
\bibitem[Casida()]{casida1995}
Casida,~M.~E. Time-Dependent Density Functional Response Theory for Molecules.
  Recent Advances in Density Functional Methods. pp 155--192\relax
\mciteBstWouldAddEndPuncttrue
\mciteSetBstMidEndSepPunct{\mcitedefaultmidpunct}
{\mcitedefaultendpunct}{\mcitedefaultseppunct}\relax
\EndOfBibitem
\bibitem[Dreuw and Head-Gordon(2005)Dreuw, and Head-Gordon]{dreuw2005}
Dreuw,~A.; Head-Gordon,~M. Single-Reference ab Initio Methods for the
  Calculation of Excited States of Large Molecules. \emph{Chem. Rev.}
  \textbf{2005}, \emph{105}, 4009\relax
\mciteBstWouldAddEndPuncttrue
\mciteSetBstMidEndSepPunct{\mcitedefaultmidpunct}
{\mcitedefaultendpunct}{\mcitedefaultseppunct}\relax
\EndOfBibitem
\bibitem[de~Silva(2019)]{desilva2019b}
de~Silva,~P. Inverted Singlet–Triplet Gaps and Their Relevance to Thermally
  Activated Delayed Fluorescence. \emph{J. Phys. Chem. Lett.} \textbf{2019},
  \emph{10}, 5674\relax
\mciteBstWouldAddEndPuncttrue
\mciteSetBstMidEndSepPunct{\mcitedefaultmidpunct}
{\mcitedefaultendpunct}{\mcitedefaultseppunct}\relax
\EndOfBibitem
\bibitem[Ghosh and Bhattacharyya(2022)Ghosh, and Bhattacharyya]{ghosh2022}
Ghosh,~S.; Bhattacharyya,~K. Origin of the Failure of Density Functional
  Theories in Predicting Inverted Singlet–Triplet Gaps. \emph{J. Phys. Chem.
  A} \textbf{2022}, \emph{126}, 1378--1385\relax
\mciteBstWouldAddEndPuncttrue
\mciteSetBstMidEndSepPunct{\mcitedefaultmidpunct}
{\mcitedefaultendpunct}{\mcitedefaultseppunct}\relax
\EndOfBibitem
\bibitem[Ricci \latin{et~al.}(2021)Ricci, San-Fabián, Olivier, and
  Sancho-García]{ricci2021}
Ricci,~G.; San-Fabián,~E.; Olivier,~Y.; Sancho-García,~J.~C. Singlet-Triplet
  Excited-State Inversion in Heptazine and Related Molecules: Assessment of
  TD-DFT and ab initio Methods. \emph{Chem. Phys. Chem.} \textbf{2021},
  \emph{22}, 553\relax
\mciteBstWouldAddEndPuncttrue
\mciteSetBstMidEndSepPunct{\mcitedefaultmidpunct}
{\mcitedefaultendpunct}{\mcitedefaultseppunct}\relax
\EndOfBibitem
\bibitem[Li \latin{et~al.}(2022)Li, Li, Liu, Gong, Zhang, Yao, and Guo]{li2022}
Li,~J.; Li,~Z.; Liu,~H.; Gong,~H.; Zhang,~J.; Yao,~Y.; Guo,~Q. Organic
  molecules with inverted singlet-triplet gaps. \emph{Front. Chem.}
  \textbf{2022}, \emph{10}\relax
\mciteBstWouldAddEndPuncttrue
\mciteSetBstMidEndSepPunct{\mcitedefaultmidpunct}
{\mcitedefaultendpunct}{\mcitedefaultseppunct}\relax
\EndOfBibitem
\bibitem[Pollice \latin{et~al.}(2021)Pollice, Friederich, Lavigne, Gomes, and
  Aspuru-Guzik]{police2021}
Pollice,~R.; Friederich,~P.; Lavigne,~C.; Gomes,~G. d.~P.; Aspuru-Guzik,~A.
  Organic Molecules with Inverted Gaps between First Excited Singlet and
  Triplet States and Appreciable Fluorescence Rates. \emph{Matter}
  \textbf{2021}, \emph{4}, 1654\relax
\mciteBstWouldAddEndPuncttrue
\mciteSetBstMidEndSepPunct{\mcitedefaultmidpunct}
{\mcitedefaultendpunct}{\mcitedefaultseppunct}\relax
\EndOfBibitem
\bibitem[Sobolewski and Domcke(2021)Sobolewski, and Domcke]{domcke2021}
Sobolewski,~A.~L.; Domcke,~W. Are Heptazine-Based Organic Light-Emitting Diode
  Chromophores Thermally Activated Delayed Fluorescence or Inverted
  Singlet-Triplet Systems? \emph{J. Phys. Chem. Lett.} \textbf{2021},
  \emph{12}, 6852\relax
\mciteBstWouldAddEndPuncttrue
\mciteSetBstMidEndSepPunct{\mcitedefaultmidpunct}
{\mcitedefaultendpunct}{\mcitedefaultseppunct}\relax
\EndOfBibitem
\bibitem[Dreuw and Hoffmann(2023)Dreuw, and Hoffmann]{dreuw2023}
Dreuw,~A.; Hoffmann,~M. The inverted singlet–triplet gap: a vanishing myth?
  \emph{Front. Chem.} \textbf{2023}, \emph{11}\relax
\mciteBstWouldAddEndPuncttrue
\mciteSetBstMidEndSepPunct{\mcitedefaultmidpunct}
{\mcitedefaultendpunct}{\mcitedefaultseppunct}\relax
\EndOfBibitem
\bibitem[Sanz-Rodrigo \latin{et~al.}(2021)Sanz-Rodrigo, Ricci, Olivier, and
  Sancho-García]{rodrigo2021}
Sanz-Rodrigo,~J.; Ricci,~G.; Olivier,~Y.; Sancho-García,~J.~C. Negative
  Singlet–Triplet Excitation Energy Gap in Triangle-Shaped Molecular Emitters
  for Efficient Triplet Harvesting. \emph{J. Phys. Chem. A} \textbf{2021},
  \emph{125}, 513\relax
\mciteBstWouldAddEndPuncttrue
\mciteSetBstMidEndSepPunct{\mcitedefaultmidpunct}
{\mcitedefaultendpunct}{\mcitedefaultseppunct}\relax
\EndOfBibitem
\bibitem[Ehrmaier \latin{et~al.}(2019)Ehrmaier, Rabe, Pristash, Corp,
  Schlenker, Sobolewski, and Domcke]{ehrmaier2019}
Ehrmaier,~J.; Rabe,~E.~J.; Pristash,~S.~R.; Corp,~K.~L.; Schlenker,~C.~W.;
  Sobolewski,~A.~L.; Domcke,~W. Singlet-Triplet Inversion in Heptazine and in
  Polymeric Carbon Nitrides. \emph{J. Phys. Chem. A} \textbf{2019}, \emph{123},
  8099\relax
\mciteBstWouldAddEndPuncttrue
\mciteSetBstMidEndSepPunct{\mcitedefaultmidpunct}
{\mcitedefaultendpunct}{\mcitedefaultseppunct}\relax
\EndOfBibitem
\bibitem[Pios \latin{et~al.}(2021)Pios, Huang, Sobolewski, and
  Domcke]{pios2021}
Pios,~S.; Huang,~X.; Sobolewski,~A.~L.; Domcke,~W. Triangular boron carbon
  nitrides: An unexplored family of chromophores with unique properties for
  photocatalysis and ptoelectronics. \emph{Phys. Chem. Chem. Phys.}
  \textbf{2021}, \emph{23}, 12968\relax
\mciteBstWouldAddEndPuncttrue
\mciteSetBstMidEndSepPunct{\mcitedefaultmidpunct}
{\mcitedefaultendpunct}{\mcitedefaultseppunct}\relax
\EndOfBibitem
\bibitem[Tučková \latin{et~al.}(2022)Tučková, Straka, Valiev, and
  Sundholm]{tuckova2022}
Tučková,~L.; Straka,~M.; Valiev,~R.~R.; Sundholm,~D. On the origin of the
  inverted singlet–triplet gap of the 5th generation light-emitting
  molecules. \emph{Phys. Chem. Chem. Phys.} \textbf{2022}, \emph{24},
  18713--18721\relax
\mciteBstWouldAddEndPuncttrue
\mciteSetBstMidEndSepPunct{\mcitedefaultmidpunct}
{\mcitedefaultendpunct}{\mcitedefaultseppunct}\relax
\EndOfBibitem
\bibitem[Loos \latin{et~al.}(2023)Loos, Lipparini, and Jacquemin]{loos2023}
Loos,~P.~F.; Lipparini,~F.; Jacquemin,~D. Heptazine, Cyclazine, and Related
  Compounds: Chemically-Accurate Estimates of the Inverted Singlet–Triplet
  Gap. \emph{The Journal of Physical Chemistry Letters} \textbf{2023},
  \emph{14}, 11069--11075\relax
\mciteBstWouldAddEndPuncttrue
\mciteSetBstMidEndSepPunct{\mcitedefaultmidpunct}
{\mcitedefaultendpunct}{\mcitedefaultseppunct}\relax
\EndOfBibitem
\bibitem[Runge and Gross(1984)Runge, and Gross]{runge1984}
Runge,~E.; Gross,~E. K.~U. Density-Functional Theory for Time Dependent
  Systems. \emph{Phys. Rev. Lett.} \textbf{1984}, \emph{52}, 997\relax
\mciteBstWouldAddEndPuncttrue
\mciteSetBstMidEndSepPunct{\mcitedefaultmidpunct}
{\mcitedefaultendpunct}{\mcitedefaultseppunct}\relax
\EndOfBibitem
\bibitem[Rhee and Head-Gordon(2007)Rhee, and Head-Gordon]{rhee2007}
Rhee,~Y.~M.; Head-Gordon,~M. Scaled Second-Order Perturbation Corrections to
  Configuration Interaction Singles: Efficient and Reliable Excitation Energy
  Methods. \emph{J. Phys. Chem. A} \textbf{2007}, \emph{111}, 5314\relax
\mciteBstWouldAddEndPuncttrue
\mciteSetBstMidEndSepPunct{\mcitedefaultmidpunct}
{\mcitedefaultendpunct}{\mcitedefaultseppunct}\relax
\EndOfBibitem
\bibitem[Head-Gordon \latin{et~al.}(1994)Head-Gordon, Rico, Oumi, and
  Lee]{headgordon1994}
Head-Gordon,~M.; Rico,~R.~J.; Oumi,~M.; Lee,~T.~J. A doubles correction to
  electronic excited states from configuration interaction in the space of
  single substitutions. \emph{Chem. Phys. Lett.} \textbf{1994}, \emph{219},
  21\relax
\mciteBstWouldAddEndPuncttrue
\mciteSetBstMidEndSepPunct{\mcitedefaultmidpunct}
{\mcitedefaultendpunct}{\mcitedefaultseppunct}\relax
\EndOfBibitem
\bibitem[Stanton and Bartlett(1993)Stanton, and Bartlett]{stanton1993}
Stanton,~J.~F.; Bartlett,~R.~J. The equation of motion coupled-cluster method.
  A systematic biorthogonal approach to molecular excitation energies,
  transition probabilities, and excited state properties. \emph{J. Chem. Phys.}
  \textbf{1993}, \emph{98}, 7029\relax
\mciteBstWouldAddEndPuncttrue
\mciteSetBstMidEndSepPunct{\mcitedefaultmidpunct}
{\mcitedefaultendpunct}{\mcitedefaultseppunct}\relax
\EndOfBibitem
\bibitem[Sancho-García \latin{et~al.}(2022)Sancho-García, Brémond, Ricci,
  Pérez-Jiménez, Olivier, and Adamo]{sanchogarchia2022}
Sancho-García,~J.~C.; Brémond,~E.; Ricci,~G.; Pérez-Jiménez,~A.~J.;
  Olivier,~Y.; Adamo,~C. {Violation of Hund’s rule in molecules: Predicting
  the excited-state energy inversion by TD-DFT with double-hybrid methods}.
  \emph{J Chem. Phys.} \textbf{2022}, \emph{156}, 034105\relax
\mciteBstWouldAddEndPuncttrue
\mciteSetBstMidEndSepPunct{\mcitedefaultmidpunct}
{\mcitedefaultendpunct}{\mcitedefaultseppunct}\relax
\EndOfBibitem
\bibitem[Hellweg \latin{et~al.}(2008)Hellweg, Grün, and Hättig]{hellweg2008}
Hellweg,~A.; Grün,~S.~A.; Hättig,~C. Benchmarking the performance of
  spin-component scaled CC2 in ground and electronically excited states.
  \emph{Phys. Chem. Chem. Phys.} \textbf{2008}, \emph{10}, 4119\relax
\mciteBstWouldAddEndPuncttrue
\mciteSetBstMidEndSepPunct{\mcitedefaultmidpunct}
{\mcitedefaultendpunct}{\mcitedefaultseppunct}\relax
\EndOfBibitem
\bibitem[Wormit \latin{et~al.}(2014)Wormit, Rehn, Harbach, Wenzel, Krauter,
  Epifanovsky, and Dreuw]{wormit2014}
Wormit,~M.; Rehn,~D.~R.; Harbach,~P. H.~P.; Wenzel,~J.; Krauter,~C.~M.;
  Epifanovsky,~E.; Dreuw,~A. Investigating excited electronic states using the
  algebraic diagrammatic construction (ADC) approach of the polarisation
  propagator. \emph{Mol. Phys.} \textbf{2014}, \emph{112}, 774\relax
\mciteBstWouldAddEndPuncttrue
\mciteSetBstMidEndSepPunct{\mcitedefaultmidpunct}
{\mcitedefaultendpunct}{\mcitedefaultseppunct}\relax
\EndOfBibitem
\bibitem[Olsen \latin{et~al.}(1988)Olsen, Roos, Jo/rgensen, and
  Jensen]{olsen1988}
Olsen,~J.; Roos,~B.~O.; Jo/rgensen,~P.; Jensen,~H. J.~A. Determinant Based
  Configuration Interaction Algorithms for Complete and Restricted
  Configuration Interaction Spaces. \emph{J. Chem. Phys.} \textbf{1988},
  \emph{89}, 2185\relax
\mciteBstWouldAddEndPuncttrue
\mciteSetBstMidEndSepPunct{\mcitedefaultmidpunct}
{\mcitedefaultendpunct}{\mcitedefaultseppunct}\relax
\EndOfBibitem
\bibitem[Angeli \latin{et~al.}(2001)Angeli, Cimiraglia, Evangelisti, Leininger,
  and Malrieu]{angeli2001}
Angeli,~C.; Cimiraglia,~R.; Evangelisti,~S.; Leininger,~T.; Malrieu,~J.-P.
  Introduction of N-electron valence states for multireference perturbation
  theory. \emph{J. Chem. Phys.} \textbf{2001}, \emph{114}, 10252\relax
\mciteBstWouldAddEndPuncttrue
\mciteSetBstMidEndSepPunct{\mcitedefaultmidpunct}
{\mcitedefaultendpunct}{\mcitedefaultseppunct}\relax
\EndOfBibitem
\bibitem[Christiansen \latin{et~al.}(1995)Christiansen, Koch, and
  Jo/rgensen]{christiansen1995}
Christiansen,~O.; Koch,~H.; Jo/rgensen,~P. {Response functions in the CC3
  iterative triple excitation model}. \emph{The Journal of Chemical Physics}
  \textbf{1995}, \emph{103}, 7429--7441\relax
\mciteBstWouldAddEndPuncttrue
\mciteSetBstMidEndSepPunct{\mcitedefaultmidpunct}
{\mcitedefaultendpunct}{\mcitedefaultseppunct}\relax
\EndOfBibitem
\bibitem[Nooijen and Bartlett(1997)Nooijen, and Bartlett]{nooijen1997}
Nooijen,~M.; Bartlett,~R.~J. Similarity transformed equation-of-motion
  coupled-cluster theory: Details, examples, and comparisons. \emph{J. Chem.
  Phys.} \textbf{1997}, \emph{107}, 6812\relax
\mciteBstWouldAddEndPuncttrue
\mciteSetBstMidEndSepPunct{\mcitedefaultmidpunct}
{\mcitedefaultendpunct}{\mcitedefaultseppunct}\relax
\EndOfBibitem
\bibitem[Ghosh \latin{et~al.}(2018)Ghosh, Verma, Cramer, Gagliardi, and
  Truhlar]{ghosh2018}
Ghosh,~S.; Verma,~P.; Cramer,~C.~J.; Gagliardi,~L.; Truhlar,~D.~G. Combining
  Wave Function Methods with Density Functional Theory for Excited States.
  \emph{Chem. Rev.} \textbf{2018}, \emph{118}, 7249\relax
\mciteBstWouldAddEndPuncttrue
\mciteSetBstMidEndSepPunct{\mcitedefaultmidpunct}
{\mcitedefaultendpunct}{\mcitedefaultseppunct}\relax
\EndOfBibitem
\bibitem[Grimme and Neese(2007)Grimme, and Neese]{grimme2007}
Grimme,~S.; Neese,~F. Double-Hybrid Density Functional Theory for Excited
  Electronic States of Molecules. \emph{J. Chem. Phys.} \textbf{2007},
  \emph{127}, 154116\relax
\mciteBstWouldAddEndPuncttrue
\mciteSetBstMidEndSepPunct{\mcitedefaultmidpunct}
{\mcitedefaultendpunct}{\mcitedefaultseppunct}\relax
\EndOfBibitem
\bibitem[Haase \latin{et~al.}(2020)Haase, Faber, Provasi, and Sauer]{haase2020}
Haase,~P. A.~B.; Faber,~R.; Provasi,~P.~F.; Sauer,~S. P.~A. Noniterative
  Doubles Corrections to the Random Phase and Higher Random Phase
  Approximations: Singlet and Triplet Excitation Energies. \emph{J Comput.
  Chem.} \textbf{2020}, \emph{41}, 43--55\relax
\mciteBstWouldAddEndPuncttrue
\mciteSetBstMidEndSepPunct{\mcitedefaultmidpunct}
{\mcitedefaultendpunct}{\mcitedefaultseppunct}\relax
\EndOfBibitem
\bibitem[Szabo and Ostlund(1996)Szabo, and Ostlund]{szabo96}
Szabo,~A.; Ostlund,~N.~S. \emph{Modern Quantum Chemistry}; {Dover}: {New York},
  1996\relax
\mciteBstWouldAddEndPuncttrue
\mciteSetBstMidEndSepPunct{\mcitedefaultmidpunct}
{\mcitedefaultendpunct}{\mcitedefaultseppunct}\relax
\EndOfBibitem
\bibitem[Helgaker \latin{et~al.}(2000)Helgaker, J{\o}rgensen, and
  Olsen]{helgaker2000}
Helgaker,~T.; J{\o}rgensen,~P.; Olsen,~J. \emph{Molecular Electronic Structure
  Theory}; {John Wiley and Sons, Ltd}: {New York}, 2000\relax
\mciteBstWouldAddEndPuncttrue
\mciteSetBstMidEndSepPunct{\mcitedefaultmidpunct}
{\mcitedefaultendpunct}{\mcitedefaultseppunct}\relax
\EndOfBibitem
\bibitem[Runge and Gross(1984)Runge, and Gross]{gross1984}
Runge,~E.; Gross,~E. K.~U. Density-Functional Theory for Time-Dependent
  Systems. \emph{Phys. Rev. Lett.} \textbf{1984}, \emph{52}, 997--1000\relax
\mciteBstWouldAddEndPuncttrue
\mciteSetBstMidEndSepPunct{\mcitedefaultmidpunct}
{\mcitedefaultendpunct}{\mcitedefaultseppunct}\relax
\EndOfBibitem
\bibitem[Burke \latin{et~al.}(2005)Burke, Werschnik, and Gross]{burke2005}
Burke,~K.; Werschnik,~J.; Gross,~E. K.~U. {Time-dependent density functional
  theory: Past, present, and future}. \emph{The Journal of Chemical Physics}
  \textbf{2005}, \emph{123}, 062206\relax
\mciteBstWouldAddEndPuncttrue
\mciteSetBstMidEndSepPunct{\mcitedefaultmidpunct}
{\mcitedefaultendpunct}{\mcitedefaultseppunct}\relax
\EndOfBibitem
\bibitem[Gross and Kohn(1985)Gross, and Kohn]{gross1985}
Gross,~E. K.~U.; Kohn,~W. Local density-functional theory of
  frequency-dependent linear response. \emph{Phys. Rev. Lett.} \textbf{1985},
  \emph{55}, 2850--2852\relax
\mciteBstWouldAddEndPuncttrue
\mciteSetBstMidEndSepPunct{\mcitedefaultmidpunct}
{\mcitedefaultendpunct}{\mcitedefaultseppunct}\relax
\EndOfBibitem
\bibitem[Burke(2012)]{burke2012}
Burke,~K. {Perspective on density functional theory}. \emph{The Journal of
  Chemical Physics} \textbf{2012}, \emph{136}, 150901\relax
\mciteBstWouldAddEndPuncttrue
\mciteSetBstMidEndSepPunct{\mcitedefaultmidpunct}
{\mcitedefaultendpunct}{\mcitedefaultseppunct}\relax
\EndOfBibitem
\bibitem[Dobson and Wang(2000)Dobson, and Wang]{dobson2000}
Dobson,~J.~F.; Wang,~J. Energy-optimized local exchange-correlation kernel for
  the electron gas: Application to van der Waals forces. \emph{Phys. Rev. B}
  \textbf{2000}, \emph{62}, 10038--10045\relax
\mciteBstWouldAddEndPuncttrue
\mciteSetBstMidEndSepPunct{\mcitedefaultmidpunct}
{\mcitedefaultendpunct}{\mcitedefaultseppunct}\relax
\EndOfBibitem
\bibitem[Elliott \latin{et~al.}(2011)Elliott, Goldson, Canahui, and
  Maitra]{elliott2011}
Elliott,~P.; Goldson,~S.; Canahui,~C.; Maitra,~N.~T. Perspectives on
  double-excitations in TDDFT. \emph{Chem. Phys.} \textbf{2011}, \emph{391},
  110\relax
\mciteBstWouldAddEndPuncttrue
\mciteSetBstMidEndSepPunct{\mcitedefaultmidpunct}
{\mcitedefaultendpunct}{\mcitedefaultseppunct}\relax
\EndOfBibitem
\bibitem[Gilbert \latin{et~al.}(2008)Gilbert, Besley, and Gill]{gilbert2008}
Gilbert,~A. T.~B.; Besley,~N.~A.; Gill,~P. M.~W. Self-Consistent Field
  Calculations of Excited States Using the Maximum Overlap Method (MOM).
  \emph{J Phys. Chem. A} \textbf{2008}, \emph{112}, 13164--13171\relax
\mciteBstWouldAddEndPuncttrue
\mciteSetBstMidEndSepPunct{\mcitedefaultmidpunct}
{\mcitedefaultendpunct}{\mcitedefaultseppunct}\relax
\EndOfBibitem
\bibitem[Hait and Head-Gordon(2020)Hait, and Head-Gordon]{hait2020}
Hait,~D.; Head-Gordon,~M. Excited State Orbital Optimization via Minimizing the
  Square of the Gradient: General Approach and Application to Singly and Doubly
  Excited States via Density Functional Theory. \emph{J Chem. Theor. Comput.}
  \textbf{2020}, \emph{16}, 1699--1710\relax
\mciteBstWouldAddEndPuncttrue
\mciteSetBstMidEndSepPunct{\mcitedefaultmidpunct}
{\mcitedefaultendpunct}{\mcitedefaultseppunct}\relax
\EndOfBibitem
\bibitem[Hait and Head-Gordon(2021)Hait, and Head-Gordon]{hait2021}
Hait,~D.; Head-Gordon,~M. Orbital optimized density functional theory for
  electronic excited states. \emph{J. Phys. Chem. Lett.} \textbf{2021},
  \emph{12}, 4517\relax
\mciteBstWouldAddEndPuncttrue
\mciteSetBstMidEndSepPunct{\mcitedefaultmidpunct}
{\mcitedefaultendpunct}{\mcitedefaultseppunct}\relax
\EndOfBibitem
\bibitem[Hait \latin{et~al.}(2016)Hait, Zhu, McMahon, and
  Van~Voorhis]{hait2016}
Hait,~D.; Zhu,~T.; McMahon,~D.~P.; Van~Voorhis,~T. Prediction of Excited-State
  Energies and Singlet–Triplet Gaps of Charge-Transfer States Using a
  Restricted Open-Shell Kohn–Sham Approach. \emph{J Chem. Theor. Comput.}
  \textbf{2016}, \emph{12}, 3353--3359\relax
\mciteBstWouldAddEndPuncttrue
\mciteSetBstMidEndSepPunct{\mcitedefaultmidpunct}
{\mcitedefaultendpunct}{\mcitedefaultseppunct}\relax
\EndOfBibitem
\bibitem[Cheng \latin{et~al.}(2008)Cheng, Wu, and Van~Voorhis]{edft2008}
Cheng,~C.-L.; Wu,~Q.; Van~Voorhis,~T. {Rydberg energies using excited state
  density functional theory}. \emph{The Journal of Chemical Physics}
  \textbf{2008}, \emph{129}, 124112\relax
\mciteBstWouldAddEndPuncttrue
\mciteSetBstMidEndSepPunct{\mcitedefaultmidpunct}
{\mcitedefaultendpunct}{\mcitedefaultseppunct}\relax
\EndOfBibitem
\bibitem[Lischka \latin{et~al.}(2018)Lischka, Nachtigallová, Aquino, Szalay,
  Plasser, Machado, and Barbatti]{lischka2018}
Lischka,~H.; Nachtigallová,~D.; Aquino,~A. J.~A.; Szalay,~P.~G.; Plasser,~F.;
  Machado,~F. B.~C.; Barbatti,~M. Multireference Approaches for Excited States
  of Molecules. \emph{Chem. Rev.} \textbf{2018}, \emph{118}, 7293--7361\relax
\mciteBstWouldAddEndPuncttrue
\mciteSetBstMidEndSepPunct{\mcitedefaultmidpunct}
{\mcitedefaultendpunct}{\mcitedefaultseppunct}\relax
\EndOfBibitem
\bibitem[Andersson \latin{et~al.}(1990)Andersson, Malmqvist, Roos, Sadlej, and
  Wolinski]{anderson1990}
Andersson,~K.; Malmqvist,~P.~A.; Roos,~B.~O.; Sadlej,~A.~J.; Wolinski,~K.
  Second-order perturbation theory with a CASSCF reference function. \emph{J.
  Phys. Chem.} \textbf{1990}, \emph{94}, 5483--5488\relax
\mciteBstWouldAddEndPuncttrue
\mciteSetBstMidEndSepPunct{\mcitedefaultmidpunct}
{\mcitedefaultendpunct}{\mcitedefaultseppunct}\relax
\EndOfBibitem
\bibitem[Andersson \latin{et~al.}(1992)Andersson, Malmqvist, and
  Roos]{Andersson1992SecondorderFunction}
Andersson,~K.; Malmqvist,~P.; Roos,~B.~O. {Second‐order perturbation theory
  with a complete active space self‐consistent field reference function}.
  \emph{J. Chem. Phys.} \textbf{1992}, \emph{96}, 1218--1226\relax
\mciteBstWouldAddEndPuncttrue
\mciteSetBstMidEndSepPunct{\mcitedefaultmidpunct}
{\mcitedefaultendpunct}{\mcitedefaultseppunct}\relax
\EndOfBibitem
\bibitem[Hirao(1992)]{Hirao1992MultireferenceMethod}
Hirao,~K. {Multireference M{\o}ller-Plesset method}. \emph{Chem. Phys. Lett.}
  \textbf{1992}, \emph{190}, 374--380\relax
\mciteBstWouldAddEndPuncttrue
\mciteSetBstMidEndSepPunct{\mcitedefaultmidpunct}
{\mcitedefaultendpunct}{\mcitedefaultseppunct}\relax
\EndOfBibitem
\bibitem[Kozlowski and Davidson(1994)Kozlowski, and
  Davidson]{Kozlowski1994ConsiderationsTheory}
Kozlowski,~P.~M.; Davidson,~E.~R. {Considerations in constructing a
  multireference second‐order perturbation theory}. \emph{J. Chem. Phys.}
  \textbf{1994}, \emph{100}, 3672--3682\relax
\mciteBstWouldAddEndPuncttrue
\mciteSetBstMidEndSepPunct{\mcitedefaultmidpunct}
{\mcitedefaultendpunct}{\mcitedefaultseppunct}\relax
\EndOfBibitem
\bibitem[Hoffmann(1996)]{Hoffmann1996CanonicalVariables}
Hoffmann,~M.~R. {Canonical Van Vleck Quasidegenerate Perturbation Theory with
  Trigonometric Variables}. \emph{J. Phys. Chem.} \textbf{1996}, \emph{100},
  6125--6130\relax
\mciteBstWouldAddEndPuncttrue
\mciteSetBstMidEndSepPunct{\mcitedefaultmidpunct}
{\mcitedefaultendpunct}{\mcitedefaultseppunct}\relax
\EndOfBibitem
\bibitem[Sen \latin{et~al.}(2015)Sen, Sen, Samanta, and
  Mukherjee]{Sen2015UnitaryApplications}
Sen,~A.; Sen,~S.; Samanta,~P.~K.; Mukherjee,~D. {Unitary group adapted state
  specific multireference perturbation theory: Formulation and pilot
  applications}. \emph{J. Comput. Chem.} \textbf{2015}, \emph{36},
  670--688\relax
\mciteBstWouldAddEndPuncttrue
\mciteSetBstMidEndSepPunct{\mcitedefaultmidpunct}
{\mcitedefaultendpunct}{\mcitedefaultseppunct}\relax
\EndOfBibitem
\bibitem[Werner and Knowles(1988)Werner, and Knowles]{Werner1988AnIN}
Werner,~H.-J.; Knowles,~P.~J. {An efficient internally contracted
  multiconfiguration-reference configuration interaction method}. \emph{J.
  Chem. Phys} \textbf{1988}, \emph{89}, 5803\relax
\mciteBstWouldAddEndPuncttrue
\mciteSetBstMidEndSepPunct{\mcitedefaultmidpunct}
{\mcitedefaultendpunct}{\mcitedefaultseppunct}\relax
\EndOfBibitem
\bibitem[Jeziorski and Monkhorst(1981)Jeziorski, and Monkhorst]{Jeziorski1981}
Jeziorski,~B.; Monkhorst,~H.~J. Coupled-cluster method for multideterminantal
  reference states. \emph{Phys. Rev. A} \textbf{1981}, \emph{24},
  1668--1681\relax
\mciteBstWouldAddEndPuncttrue
\mciteSetBstMidEndSepPunct{\mcitedefaultmidpunct}
{\mcitedefaultendpunct}{\mcitedefaultseppunct}\relax
\EndOfBibitem
\bibitem[Li and Paldus(1995)Li, and Paldus]{Li1995UnitaryApproximations}
Li,~X.; Paldus,~J. {Unitary group based state‐selective coupled‐cluster
  method: Comparison of the first order interacting space and the full single
  and double excitation space approximations}. \emph{J. Chem. Phys.}
  \textbf{1995}, \emph{102}, 8897--8905\relax
\mciteBstWouldAddEndPuncttrue
\mciteSetBstMidEndSepPunct{\mcitedefaultmidpunct}
{\mcitedefaultendpunct}{\mcitedefaultseppunct}\relax
\EndOfBibitem
\bibitem[Li and Paldus(1997)Li, and Paldus]{Li1997ReducedStates}
Li,~X.; Paldus,~J. {Reduced multireference CCSD method: An effective approach
  to quasidegenerate states}. \emph{J. Chem. Phys.} \textbf{1997}, \emph{107},
  6257--6269\relax
\mciteBstWouldAddEndPuncttrue
\mciteSetBstMidEndSepPunct{\mcitedefaultmidpunct}
{\mcitedefaultendpunct}{\mcitedefaultseppunct}\relax
\EndOfBibitem
\bibitem[Mahapatra \latin{et~al.}(1998)Mahapatra, Datta, and
  Mukherjee]{MAHAPATRA1998AApplications}
Mahapatra,~U.~S.; Datta,~B.; Mukherjee,~D. {A state-specific multi-reference
  coupled cluster formalism with molecular applications}. \emph{Mol. Phys.}
  \textbf{1998}, \emph{94}, 157--171\relax
\mciteBstWouldAddEndPuncttrue
\mciteSetBstMidEndSepPunct{\mcitedefaultmidpunct}
{\mcitedefaultendpunct}{\mcitedefaultseppunct}\relax
\EndOfBibitem
\bibitem[Mahapatra \latin{et~al.}(1998)Mahapatra, Datta, Bandyopadhyay, and
  Mukherjee]{MAHAPATRA1998163}
Mahapatra,~U.~S.; Datta,~B.; Bandyopadhyay,~B.; Mukherjee,~D. In
  \emph{State-Specific Multi-Reference Coupled Cluster Formulations: Two
  Paradigms}; Löwdin,~P.-O., Ed.; Advances in Quantum Chemistry; Academic
  Press, 1998; Vol.~30; pp 163--193\relax
\mciteBstWouldAddEndPuncttrue
\mciteSetBstMidEndSepPunct{\mcitedefaultmidpunct}
{\mcitedefaultendpunct}{\mcitedefaultseppunct}\relax
\EndOfBibitem
\bibitem[Aoto and Köhn(2016)Aoto, and Köhn]{aoto2016}
Aoto,~Y.~A.; Köhn,~A. {Internally contracted multireference coupled-cluster
  theory in a multistate framework}. \emph{J. Chem. Phys.} \textbf{2016},
  \emph{144}, 074103\relax
\mciteBstWouldAddEndPuncttrue
\mciteSetBstMidEndSepPunct{\mcitedefaultmidpunct}
{\mcitedefaultendpunct}{\mcitedefaultseppunct}\relax
\EndOfBibitem
\bibitem[Szalay and Bartlett(1995)Szalay, and Bartlett]{Szalay1995}
Szalay,~P.~G.; Bartlett,~R.~J. {Approximately extensive modifications of the
  multireference configuration interaction method: A theoretical and practical
  analysis}. \emph{J. Chem. Phys.} \textbf{1995}, \emph{103}, 3600--3612\relax
\mciteBstWouldAddEndPuncttrue
\mciteSetBstMidEndSepPunct{\mcitedefaultmidpunct}
{\mcitedefaultendpunct}{\mcitedefaultseppunct}\relax
\EndOfBibitem
\bibitem[Neese \latin{et~al.}(2020)Neese, Wennmohs, Becker, and
  Riplinger]{orca2020}
Neese,~F.; Wennmohs,~F.; Becker,~U.; Riplinger,~C. {The ORCA quantum chemistry
  program package}. \emph{J. Chem. Phys.} \textbf{2020}, \emph{152},
  224108\relax
\mciteBstWouldAddEndPuncttrue
\mciteSetBstMidEndSepPunct{\mcitedefaultmidpunct}
{\mcitedefaultendpunct}{\mcitedefaultseppunct}\relax
\EndOfBibitem
\bibitem[Shao \latin{et~al.}(2015)Shao, Gan, Epifanovsky, Gilbert, Wormit,
  Kussmann, Lange, Behn, Deng, and Feng]{shao2015}
Shao,~Y.; Gan,~Z.; Epifanovsky,~E.; Gilbert,~A.~T.; Wormit,~M.; Kussmann,~J.;
  Lange,~A.~W.; Behn,~A.; Deng,~J.; Feng,~X. Advances in molecular quantum
  chemistry contained in the Q-Chem 4 program package. \emph{Mol. Phys.}
  \textbf{2015}, \emph{113}, 184\relax
\mciteBstWouldAddEndPuncttrue
\mciteSetBstMidEndSepPunct{\mcitedefaultmidpunct}
{\mcitedefaultendpunct}{\mcitedefaultseppunct}\relax
\EndOfBibitem
\bibitem[Weigend and Ahlrichs(2005)Weigend, and Ahlrichs]{weigend2005}
Weigend,~F.; Ahlrichs,~R. Balanced basis sets of split valence, triple zeta
  valence and quadruple zeta valence quality for H to Rn: Design and assessment
  of accuracy. \emph{Phys. Chem. Chem. Phys.} \textbf{2005}, \emph{7},
  3297\relax
\mciteBstWouldAddEndPuncttrue
\mciteSetBstMidEndSepPunct{\mcitedefaultmidpunct}
{\mcitedefaultendpunct}{\mcitedefaultseppunct}\relax
\EndOfBibitem
\bibitem[Stanton(1997)]{Stanton1997}
Stanton,~J.~F. Why CCSD(T) works: a different perspective. \emph{Chem. Phys.
  Lett.} \textbf{1997}, \emph{281}, 130--134\relax
\mciteBstWouldAddEndPuncttrue
\mciteSetBstMidEndSepPunct{\mcitedefaultmidpunct}
{\mcitedefaultendpunct}{\mcitedefaultseppunct}\relax
\EndOfBibitem
\bibitem[Raghavachari \latin{et~al.}(1989)Raghavachari, Trucks, Pople, and
  Head-Gordon]{Raghavachari1989}
Raghavachari,~K.; Trucks,~G.~W.; Pople,~J.~A.; Head-Gordon,~M. A fifth-order
  perturbation comparison of electron correlation theories. \emph{Chem. Phys.
  Lett.} \textbf{1989}, \emph{157}, 479--483\relax
\mciteBstWouldAddEndPuncttrue
\mciteSetBstMidEndSepPunct{\mcitedefaultmidpunct}
{\mcitedefaultendpunct}{\mcitedefaultseppunct}\relax
\EndOfBibitem
\bibitem[Dunning(1989)]{dunning1989}
Dunning,~J.,~Thom~H. {Gaussian basis sets for use in correlated molecular
  calculations. I. The atoms boron through neon and hydrogen}. \emph{J. Chem.
  Phys.} \textbf{1989}, \emph{90}, 1007--1023\relax
\mciteBstWouldAddEndPuncttrue
\mciteSetBstMidEndSepPunct{\mcitedefaultmidpunct}
{\mcitedefaultendpunct}{\mcitedefaultseppunct}\relax
\EndOfBibitem
\bibitem[Becke(1993)]{becke1993}
Becke,~A.~D. Density-functional thermochemistry. III. The role of exact
  exchange. \emph{J. Chem. Phys.} \textbf{1993}, \emph{98}, 5648\relax
\mciteBstWouldAddEndPuncttrue
\mciteSetBstMidEndSepPunct{\mcitedefaultmidpunct}
{\mcitedefaultendpunct}{\mcitedefaultseppunct}\relax
\EndOfBibitem
\bibitem[Grimme \latin{et~al.}(2011)Grimme, Ehrlich, and Goerigk]{grimme2011}
Grimme,~S.; Ehrlich,~S.; Goerigk,~L. Effect of the damping function in
  dispersion corrected density functional theory. \emph{J. Comput. Chem.}
  \textbf{2011}, \emph{32}, 1456--1465\relax
\mciteBstWouldAddEndPuncttrue
\mciteSetBstMidEndSepPunct{\mcitedefaultmidpunct}
{\mcitedefaultendpunct}{\mcitedefaultseppunct}\relax
\EndOfBibitem
\bibitem[Bhattacharyya(2021)]{bhattacharyya2021}
Bhattacharyya,~K. Can TDDFT Render the Electronic Excited States Ordering of
  Azine Derivative? A Closer Investigation with DLPNO-STEOM-CCSD. \emph{Chem.
  Phys. Lett.} \textbf{2021}, \emph{779}, 138827\relax
\mciteBstWouldAddEndPuncttrue
\mciteSetBstMidEndSepPunct{\mcitedefaultmidpunct}
{\mcitedefaultendpunct}{\mcitedefaultseppunct}\relax
\EndOfBibitem
\bibitem[Stanton(1994)]{Stanton1994a}
Stanton,~J.~F. On the extent of spin contamination in open‐shell
  coupled‐cluster wave functions. \emph{J Chem. Phys.} \textbf{1994},
  \emph{101}, 371--374\relax
\mciteBstWouldAddEndPuncttrue
\mciteSetBstMidEndSepPunct{\mcitedefaultmidpunct}
{\mcitedefaultendpunct}{\mcitedefaultseppunct}\relax
\EndOfBibitem
\bibitem[Drwal \latin{et~al.}(2023)Drwal, Matousek, Golub, Tucholska, Hapka,
  Brabec, Veis, and Pernal]{drwal2023}
Drwal,~D.; Matousek,~M.; Golub,~P.; Tucholska,~A.; Hapka,~M.; Brabec,~J.;
  Veis,~L.; Pernal,~K. Role of Spin Polarization and Dynamic Correlation in
  Singlet–Triplet Gap Inversion of Heptazine Derivatives. \emph{J Chem.
  Theor. Comput.} \textbf{2023}, \emph{19}, 7606--7616\relax
\mciteBstWouldAddEndPuncttrue
\mciteSetBstMidEndSepPunct{\mcitedefaultmidpunct}
{\mcitedefaultendpunct}{\mcitedefaultseppunct}\relax
\EndOfBibitem
\bibitem[Ye and Voorhis(2020)Ye, and Voorhis]{ye2020arxiv}
Ye,~H.~Z.; Voorhis,~T.~V. Self-consistent M{\o}ller-Plesset Perturbation Theory
  For Excited States. \emph{arXiv preprint~arXiv:~2008.10777} \textbf{2020},
  \relax
\mciteBstWouldAddEndPunctfalse
\mciteSetBstMidEndSepPunct{\mcitedefaultmidpunct}
{}{\mcitedefaultseppunct}\relax
\EndOfBibitem
\bibitem[Rettig \latin{et~al.}(2020)Rettig, Hait, Bertels, and
  Head-Gordon]{rettig2020}
Rettig,~A.; Hait,~D.; Bertels,~L.~W.; Head-Gordon,~M. Third-Order
  Møller–Plesset Theory Made More Useful? The Role of Density Functional
  Theory Orbitals. \emph{Journal of Chemical Theory and Computation}
  \textbf{2020}, \emph{16}, 7473--7489\relax
\mciteBstWouldAddEndPuncttrue
\mciteSetBstMidEndSepPunct{\mcitedefaultmidpunct}
{\mcitedefaultendpunct}{\mcitedefaultseppunct}\relax
\EndOfBibitem
\bibitem[Fang \latin{et~al.}(2016)Fang, Lee, Peterson, and Dixon]{fang2016}
Fang,~Z.; Lee,~Z.; Peterson,~K.~A.; Dixon,~D.~A. Use of Improved Orbitals for
  CCSD(T) Calculations for Predicting Heats of Formation of Group IV and Group
  VI Metal Oxide Monomers and Dimers and UCl6. \emph{Journal of Chemical Theory
  and Computation} \textbf{2016}, \emph{12}, 3583--3592\relax
\mciteBstWouldAddEndPuncttrue
\mciteSetBstMidEndSepPunct{\mcitedefaultmidpunct}
{\mcitedefaultendpunct}{\mcitedefaultseppunct}\relax
\EndOfBibitem
\bibitem[Beran \latin{et~al.}(2003)Beran, Gwaltney, and Head-Gordon]{beran2003}
Beran,~G. J.~O.; Gwaltney,~S.~R.; Head-Gordon,~M. Approaching closed-shell
  accuracy for radicals using coupled cluster theory with perturbative triple
  substitutions. \emph{Phys. Chem. Chem. Phys.} \textbf{2003}, \emph{5},
  2488--2493\relax
\mciteBstWouldAddEndPuncttrue
\mciteSetBstMidEndSepPunct{\mcitedefaultmidpunct}
{\mcitedefaultendpunct}{\mcitedefaultseppunct}\relax
\EndOfBibitem
\bibitem[Mallick \latin{et~al.}(2021)Mallick, Rai, and
  Kumar]{MALLICK2021113326}
Mallick,~S.; Rai,~P.~K.; Kumar,~P. Accurate estimation of singlet-triplet gap
  of strongly correlated systems by CCSD(T) method using improved orbitals.
  \emph{Comput. Theor. Chem.} \textbf{2021}, \emph{1202}, 113326\relax
\mciteBstWouldAddEndPuncttrue
\mciteSetBstMidEndSepPunct{\mcitedefaultmidpunct}
{\mcitedefaultendpunct}{\mcitedefaultseppunct}\relax
\EndOfBibitem
\bibitem[Alipour and Izadkhast(2022)Alipour, and Izadkhast]{mojtaba2022}
Alipour,~M.; Izadkhast,~T. {Do any types of double-hybrid models render the
  correct order of excited state energies in inverted singlet–triplet
  emitters?} \emph{J Chem. Phys.} \textbf{2022}, \emph{156}, 064302\relax
\mciteBstWouldAddEndPuncttrue
\mciteSetBstMidEndSepPunct{\mcitedefaultmidpunct}
{\mcitedefaultendpunct}{\mcitedefaultseppunct}\relax
\EndOfBibitem
\bibitem[JAN(1998)]{JANSSEN1998423}
New diagnostics for coupled-cluster and Møller–Plesset perturbation theory.
  \emph{Chemical Physics Letters} \textbf{1998}, \emph{290}, 423--430\relax
\mciteBstWouldAddEndPuncttrue
\mciteSetBstMidEndSepPunct{\mcitedefaultmidpunct}
{\mcitedefaultendpunct}{\mcitedefaultseppunct}\relax
\EndOfBibitem
\bibitem[Lee and Taylor(1989)Lee, and Taylor]{lee1989}
Lee,~T.~J.; Taylor,~P.~R. A diagnostic for determining the quality of
  single-reference electron correlation methods. \emph{International Journal of
  Quantum Chemistry} \textbf{1989}, \emph{36}, 199--207\relax
\mciteBstWouldAddEndPuncttrue
\mciteSetBstMidEndSepPunct{\mcitedefaultmidpunct}
{\mcitedefaultendpunct}{\mcitedefaultseppunct}\relax
\EndOfBibitem
\bibitem[Bao and Truhlar(2019)Bao, and Truhlar]{bao2019}
Bao,~J.~J.; Truhlar,~D.~G. Automatic Active Space Selection for Calculating
  Electronic Excitation Energies Based on High-Spin Unrestricted Hartree–Fock
  Orbitals. \emph{J Chem. Theor. Comput.} \textbf{2019}, \emph{15},
  5308--5318\relax
\mciteBstWouldAddEndPuncttrue
\mciteSetBstMidEndSepPunct{\mcitedefaultmidpunct}
{\mcitedefaultendpunct}{\mcitedefaultseppunct}\relax
\EndOfBibitem
\bibitem[Golub \latin{et~al.}(2021)Golub, Antalik, Veis, and Brabec]{golub2021}
Golub,~P.; Antalik,~A.; Veis,~L.; Brabec,~J. Machine Learning-Assisted
  Selection of Active Spaces for Strongly Correlated Transition Metal Systems.
  \emph{J Chem. Theor. Comput.} \textbf{2021}, \emph{17}, 6053--6072\relax
\mciteBstWouldAddEndPuncttrue
\mciteSetBstMidEndSepPunct{\mcitedefaultmidpunct}
{\mcitedefaultendpunct}{\mcitedefaultseppunct}\relax
\EndOfBibitem
\bibitem[Stein and Reiher(2019)Stein, and Reiher]{stein2019}
Stein,~C.~J.; Reiher,~M. autoCAS: A Program for Fully Automated
  Multiconfigurational Calculations. \emph{J. Comput. Chem.} \textbf{2019},
  \emph{40}, 2216--2226\relax
\mciteBstWouldAddEndPuncttrue
\mciteSetBstMidEndSepPunct{\mcitedefaultmidpunct}
{\mcitedefaultendpunct}{\mcitedefaultseppunct}\relax
\EndOfBibitem
\bibitem[King and Gagliardi(2021)King, and Gagliardi]{king2021}
King,~D.~S.; Gagliardi,~L. A Ranked-Orbital Approach to Select Active Spaces
  for High-Throughput Multireference Computation. \emph{J Chem. Theor. Comput.}
  \textbf{2021}, \emph{17}, 2817--2831\relax
\mciteBstWouldAddEndPuncttrue
\mciteSetBstMidEndSepPunct{\mcitedefaultmidpunct}
{\mcitedefaultendpunct}{\mcitedefaultseppunct}\relax
\EndOfBibitem
\bibitem[Veryazov \latin{et~al.}(2011)Veryazov, Malmqvist, and
  Roos]{veryazov2021}
Veryazov,~V.; Malmqvist,~P.~A.; Roos,~B.~O. How to select active space for
  multiconfigurational quantum chemistry? \emph{Int. J Quant. Chem.}
  \textbf{2011}, \emph{111}, 3329--3338\relax
\mciteBstWouldAddEndPuncttrue
\mciteSetBstMidEndSepPunct{\mcitedefaultmidpunct}
{\mcitedefaultendpunct}{\mcitedefaultseppunct}\relax
\EndOfBibitem
\bibitem[Loos \latin{et~al.}(2019)Loos, Boggio~Pasqua, Scemama, Caffarel, and
  Jacquemin]{loos2019}
Loos,~P.~F.; Boggio~Pasqua,~M.; Scemama,~A.; Caffarel,~M.; Jacquemin,~D.
  Reference energies for double excitations. \emph{J. Chem. Theory Comput.}
  \textbf{2019}, \emph{15}, 1939\relax
\mciteBstWouldAddEndPuncttrue
\mciteSetBstMidEndSepPunct{\mcitedefaultmidpunct}
{\mcitedefaultendpunct}{\mcitedefaultseppunct}\relax
\EndOfBibitem
\end{mcitethebibliography}

\newpage

\begin{figure}[h]
\centering
	\includegraphics[width=\textwidth]{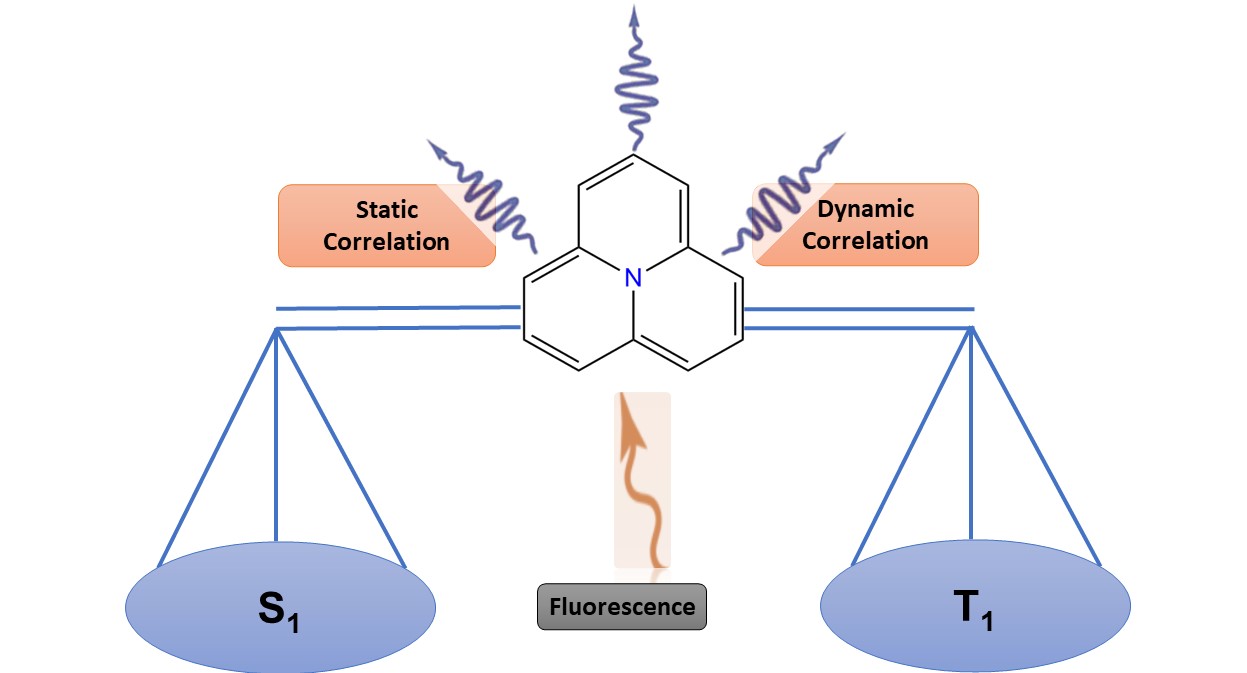}
\end{figure}

\end{document}